\documentclass[fleqn,usenatbib]{mnras}

\usepackage{newtxtext,newtxmath}

\usepackage[T1]{fontenc}
\usepackage{ae,aecompl}
\usepackage{xcolor}
\usepackage{caption}
\usepackage{subcaption}
\usepackage{hyperref}

\usepackage{graphicx}	
\usepackage{amsmath}	

\usepackage{comment}

\newcommand{\jarm}[1]{\textcolor{black}{#1}}
\newcommand{\rtgs}[1]{\textcolor{black}{#1}}
\newcommand{\chlc}[1]{\textcolor{black}{#1}}

\title[QUIJOTE MFI wide survey radio sources]{QUIJOTE scientific results -- IX. 
Radio sources in the QUIJOTE-MFI wide survey maps}

\author[D. Herranz et al.]{D. Herranz,$^1$\thanks{e-mail:herranz@ifca.unican.es} 
M. L\'opez-Caniego,$^{2,3}$
C.~H. L\'opez-Caraballo,$^{4,5}$
R.~T. G\'enova-Santos,$^{4,5}$
\newauthor
Y.~C. Perrott,$^{6}$
J.~A. Rubi{\~n}o-Mart\'{\i}n,$^{4,5}$
R. Rebolo,$^{4,5,7}$
E. Artal,$^{8}$
M. Ashdown,$^{9,10}$
\newauthor
R.~B. Barreiro,$^{1}$
F.~J. Casas,$^{1}$
E. de la Hoz,$^{1,11}$
M. Fern\'andez-Torreiro,$^{4,5}$
F. Guidi,$^{4,5,12}$
\newauthor
R.~J. Hoyland,$^{4,5}$
A.~N. Lasenby,$^{9,10}$
E. Mart\'\i nez-Gonz\'alez,$^{1}$
M.~W. Peel,$^{4,5}$ 
\newauthor
L. Piccirillo,$^{13}$
F. Poidevin,$^{4,5}$
B. Ruiz-Granados,$^{4,5,14}$
D. Tramonte,$^{15,16,4,5}$
\newauthor
F. Vansyngel,$^{4,5}$
P. Vielva,$^{1}$
R.~A. Watson.$^{13}$ 
\\
 \\
$^{1}$Instituto de F\'isica de Cantabria (IFCA), CSIC-Univ. de Cantabria, Avda. los Castros s/n, E-39005 Santander, Spain.\\
$^{2}$Aurora Technology for the European Space Agency (ESA), European Space Astronomy Centre (ESAC),  Camino Bajo del Castillo s/n, 28692 \\Villanueva de la Ca\~nada, Madrid, Spain. \\
$^{3}$Universidad Europea de Madrid, 28670, Madrid, Spain. \\
$^{4}$Instituto de Astrof\'{\i}sica de Canarias, E-38200 La Laguna, Tenerife, Spain.\\
$^{5}$Departamento de Astrof\'{\i}sica, Universidad de La Laguna, E-38206 La Laguna, Tenerife, Spain.\\
$^{6}$School of Chemical and Physical Sciences, Victoria University of Wellington, PO Box 600, Wellington 6140, New Zealand. \\
$^{7}$Consejo Superior de Investigaciones Cientificas, E-28006 Madrid, Spain \\ 
$^{8}$Departamento de Ingenieria de COMunicaciones (DICOM), Edificio Ingenieria de Telecomunicacion, Plaza de la Ciencia s/n, \\
E-39005 Santander, Spain.\\
$^{9}$Astrophysics Group, Cavendish Laboratory, University of Cambridge, J J Thomson Avenue, Cambridge CB3 0HE, U.K.\\
$^{10}$Kavli Institute for Cosmology, University of Cambridge, Madingley Road, Cambridge CB3 0HA, U.K.\\
$^{11}$Dpto. de F\'isica Moderna, Universidad de Cantabria, Avda. de los Castros s/n, E-39005 Santander, Spain.  \\
$^{12}$Institut d'Astrophysique de Paris, UMR 7095, CNRS \& Sorbonne Universit\'e, 98 bis boulevard Arago, 75014 Paris, France.\\
$^{13}$Jodrell Bank Centre for Astrophysics, Alan Turing Building, Department of Physics and Astronomy, School of Nature Sciences, \\
University of Manchester, Oxford Road, Manchester M13 9PL, U.K.\\
$^{14}$Departamento de F\'{\i}sica. Facultad de Ciencias. Universidad de C\'ordoba.  Campus de Rabanales, 
Edif. C2. Planta Baja.  E-14071 \\
C\'ordoba, Spain.\\
$^{15}$Purple Mountain Observatory, CAS, No.10 Yuanhua Road, Qixia District, Nanjing 210034, China. \\
$^{16}$NAOC-UKZN Computational Astrophysics Center (NUCAC), University of Kwazulu-Natal, Durban 4000, South Africa. 
}

\date{Accepted XXX. Received YYY; in original form ZZZ.}

\pubyear{2022}

\begin{document}
\label{firstpage}
\pagerange{\pageref{firstpage}--\pageref{lastpage}}
\maketitle

\begin{abstract}
We present the catalogue of  Q-U-I JOint TEnerife (QUIJOTE) Wide Survey radio 
sources extracted from the maps of the Multi-Frequency Instrument compiled between 2012 
and 2018. The catalogue contains 786 sources observed in intensity and  polarization, 
and is divided into two separate sub-catalogues: one containing 47 bright sources 
previously studied by the \emph{Planck} collaboration and an extended catalogue 
of 739 sources either selected from the \emph{Planck} Second Catalogue of Compact Sources or 
found through a blind search carried out with a Mexican Hat 2 wavelet. A significant 
fraction  of the sources in our catalogue (38.7 per cent) are within the
 $|b| \leq 20^\circ$ region of the Galactic plane. 
We determine statistical properties
for those sources that are likely to be extragalactic. We find that these statistical
 properties are compatible with currently available models, with a $\sim$1.8 Jy completeness 
limit at 11 GHz. We provide the polarimetric properties of (38, 33, 31, 23) 
sources with P detected above the $99.99\%$ significance level at (11, 13, 17, 19) GHz 
respectively.  Median polarization fractions are in the $2.8$--$4.7$\% range in 
the 11--19 GHz frequency interval. We do not distinguish between Galactic and extragalactic 
sources here. The results presented here are consistent with those reported in the 
literature for flat- and  steep-spectrum radio sources. 
\end{abstract}


\section{Introduction}

Cosmic Microwave Background (CMB) surveys provide
not only a wealth of information about cosmological 
parameters and the conditions of the Universe near the 
epoch of recombination, but also a probe of Galactic 
and extragalactic foregrounds; that is, astrophysical 
processes of CMB photons along the line of sight. In 
particular, CMB experiments are sensitive to 
 samples of extragalactic radio sources at much higher 
frequencies than traditional radio surveys.
CMB surveys of extragalactic sources 
allow us to investigate the evolutionary properties of 
blazar populations, to study the earliest and latest 
stages of radio activity in galaxies, and to discover 
new phenomena including new transient sources and events
\citep[see, for example,][]{review_deZotti,zotti19}. 
Additionally, CMB experiments are providing some of the 
first direct
  source number counts in polarization above $\sim$10 GHz 
\citep{QUIET,PCCS2,bonavera17,ACTpol} that, together with other 
specific ground-based focused surveys 
\citep{jackson10,VSA10,Battye11,sajina11,galluzzi16,PACO18}, 
are enabling for the first time a solid assessment to be made of the 
extragalactic source contamination of CMB maps. Moreover, 
this allows us to
    better understand the structure and intensity of 
magnetic fields, particle densities and structures of 
emitting regions inside radio sources.
 
 Even if we were interested only in early-universe cosmology 
and not  in extragalactic sources per se, it is 
still necessary  
 either to subtract or mask the contamination due to point 
sources before addressing the statistical analysis of the 
CMB. As discussed by \cite{Tucci05,Battye11,Tucci12,Puglisi}, 
polarized point sources 
 constitute a dominant foreground that
 can contaminate the cosmological B-mode polarization if 
the tensor-to-scalar ratio is $< 0.05$, and they have to 
be robustly controlled to de-lens CMB B-modes on an angular scale of arc minutes
 \citep[for the importance of de-lensing CMB B-modes see, 
for example,][]{cmblensing}. The importance, therefore, of 
detecting and characterizing the population of polarized 
point sources in the frequency range $\nu \geq 10$\, GHz 
cannot be overlooked.

The QUIJOTE (Q-U-I JOint TEnerife) experiment\footnote{QUIJOTE 
web page: \url{http://research.iac.es/proyecto/cmb/pages/en/quijote-cmb-experiment.php}} \citep{Rubino10}
is a scientific collaboration between the Instituto de 
Astrof\'\i sica de Canarias, the Instituto de F\'\i sica 
de Cantabria, the universities of Cantabria, Manchester and 
Cambridge, and the IDOM company. 
QUIJOTE is a 
polarimeter with the task of characterizing the polarization 
of the Cosmic Microwave Background, and other Galactic or 
extragalactic physical processes, including Galactic and 
extragalactic point sources that emit in microwaves in the 
frequency range 10--40 GHz. The experiment has been designed 
to reach the required sensitivity to detect a primordial 
gravitational wave component in the CMB, provided its 
tensor-to-scalar ratio is greater than $r \sim 0.05$, but to 
reach this goal it is necessary to  
characterize in detail the physical properties of the principal radio 
foregrounds (again, including point sources) in the frequency 
range covered by the experiment.
The project consists of two telescopes equipped with three 
instruments: the Multi-Frequency Instrument (hereafter, MFI), 
operating at 10--20GHz, the Thirty-GHz Instrument (TGI) and 
the Forty-GHz Instrument (FGI).

The MFI
 consists of four horns that provide eight independent output 
channels in the range 10--20\,GHz. Horns 1 and 3 contain 
10--14\,GHz polarimeters, and horns 2 and 4 contain 
16--20\,GHz polarimeters. Using frequency filters in the 
back-end of the instrument, each horn provides 
two output frequency channels, each one with a bandwidth
of approximately $\Delta \nu = 2$\,GHz. In total, the MFI provides four frequency bands
centred around 11, 13, 17 and 19\,GHz, each band being covered by two
independent horns. 
Although wide-survey maps made by horn 1
have been produced for internal consistency tests, they have not
been used for this paper because they are significantly affected by
systematic effects \citep[see][for more details]{mfiwidesurvey}.
The approximate angular resolution, given in terms of 
the full width at
half maximum, is 55\,arcmin for the low-frequency bands at 
11 and 13\,GHz, and 40\,arcmin for the 17 and 19\,GHz channels.
The technical details of the MFI are described in detail 
in \cite{MFIstatus12}. A thorough description of the MFI data 
processing pipeline can be found in \cite{MFIpipeline}. 

The QUIJOTE Wide Survey is a shallow survey that covers all 
the sky visible from Teide Observatory with elevations 
greater than $30^\circ$. This was one of the main scientific 
objectives of QUIJOTE \citep{RubinoSPIE12} and
of the MFI instrument in particular, which was in operation between 2012 and 2018. 
A detailed description of the Wide Survey and its scanning 
strategy, sky coverage and maps can be found 
in \cite{mfiwidesurvey}. 
The QUIJOTE Wide Survey provides a unique view of a large 
portion of the sky at a frequency range of utmost importance 
for the characterization of radio foregrounds for CMB science. 
The final sensitivity of the QUIJOTE MFI Wide Survey maps is 
in the range 65--200~$\mu$K per 1-degree beam in total intensity 
and 35--40~$\mu$K per 1-degree beam in polarization, depending 
on the horn and frequency, with the low-frequency channels 
having better sensitivity thanks to lower atmospheric noise. 
In this paper we focus on the identification and study of a 
comprehensive catalogue of compact  radio sources observed both 
in temperature and polarization in the Wide Survey. 
Although the angular resolution of QUIJOTE is not ideal for point 
source studies (it was not designed for that purpose), it is still interesting to 
exploit 
 the data and instrument capabilities to the full in order to make 
this study a part of the effort to characterize the radio 
foregrounds in this range of frequencies.

The structure of this paper is as follows. 
In Section~\ref{sec:source_samples} we describe the samples 
of radio sources that we shall study in this paper. 
In Section~\ref{sec:FF} we review the method that we use to 
study the linear polarization properties of our sample of radio 
sources. The main products are presented in Section~\ref{sec:products}, 
where we describe the catalogue of radio sources in intensity and polarization, and validate
its internal consistency. In section~\ref{sec:analysis} we 
 perform a brief statistical analysis of the sources detected 
with the QUIJOTE-MFI Wide Survey between 11 and 19 GHz, and we 
describe the properties of the sources in the catalogue in both 
temperature and polarization. 
 In Section~\ref{sec:variability} we study the variability of 
the sources that have been detected with high signal-to-noise 
ratio (SNR $\geq 5$) at 11 GHz.
 Finally, in Section~\ref{sec:discussion} we discuss the results 
and give our conclusions.

\section{Point source samples} \label{sec:source_samples}

In this paper we  study the polarimetric properties of a 
total intensity-selected sample of point sources located in 
the QUIJOTE Wide Survey footprint. We  follow a double sample 
selection strategy. On the one hand, we  study a non-blind 
sample of bright sources previously studied at higher 
observing frequencies. This non-blind sample  allows us to 
focus on interesting sources that might (or might not) be missed 
by a blind search because of the angular resolution and sensitivity 
of QUIJOTE. We also perform a blind search in 
the  QUIJOTE Wide Survey temperature maps. With this blind 
sample we  try to find
objects (mainly steep-spectrum radio sources)  that are 
observable at QUIJOTE MFI frequencies but are not included 
in the \emph{Planck} source catalogues. As a flux-limited search, 
the blind sample will also be suitable for statistical analyses 
such as completeness limits, number counts, etc.

\subsection{Non-blind input sample} \label{sec:non_blind}

We have selected two non-blind input samples for our study. 
The main sample (MS) consists of bright radio sources whose 
polarization has been measured at 30 GHz with statistical 
significance equal to or greater than $99.99\%$ in the \textit{Planck} 
Second Catalogue of Compact Sources \citep[PCCS2,][]{PCCS2}. 
The PCCS2 contains 114 sources with polarization detected above
the $99.99\%$ confidence limit. 
Among these 114 sources, 47 lie 
outside the region masked by this study. 
This mask is the
 QUIJOTE Wide Survey 
 sat+NCP+lowdec
mask proposed for analysis in Section 3.1 of \cite{mfiwidesurvey} 
that covers the unobserved and contaminated regions of the sky,
 and is represented in Figure~\ref{fig:figs_one_to_three} by the grey area.
In 
order to minimize possible border effects during the filtering 
process to be described in Section~\ref{sec:FF}, we extend the 
masking in the following way:  we draw a five-degree radius
 circle around each target position. If such a circle has
more than a quarter of the pixels
excluded by the mask, we remove  that position from our input catalogue.
Out of the 47 sources that remain in our main sample, 33 lie near 
the Galactic plane   band $|b|\leq 20^\circ$ band. This sample includes
 well-known sources such as Tau A, Virgo A, and radio sources 3C273,
       3C286,
       3C405 and
       3C461.
Figure~\ref{fig:figs_one_to_three} shows, in the top panel, the 
positions on the sky, in Galactic coordinates, of the 47 sources 
in our input \textit{Planck} sample. This main sample provides the 
opportunity to check our polarimetry against \textit{Planck} values 
(taking into consideration the differences in frequency and epoch of 
observation) and, where possible, to study the spectral energy 
distribution (SED) in both intensity and polarization of these 
sources between 11 and 30 GHz and/or variability of these sources.  

\begin{figure} 
     \centering
     \begin{subfigure}[b]{\columnwidth}
         \centering
         \includegraphics[width=\columnwidth]{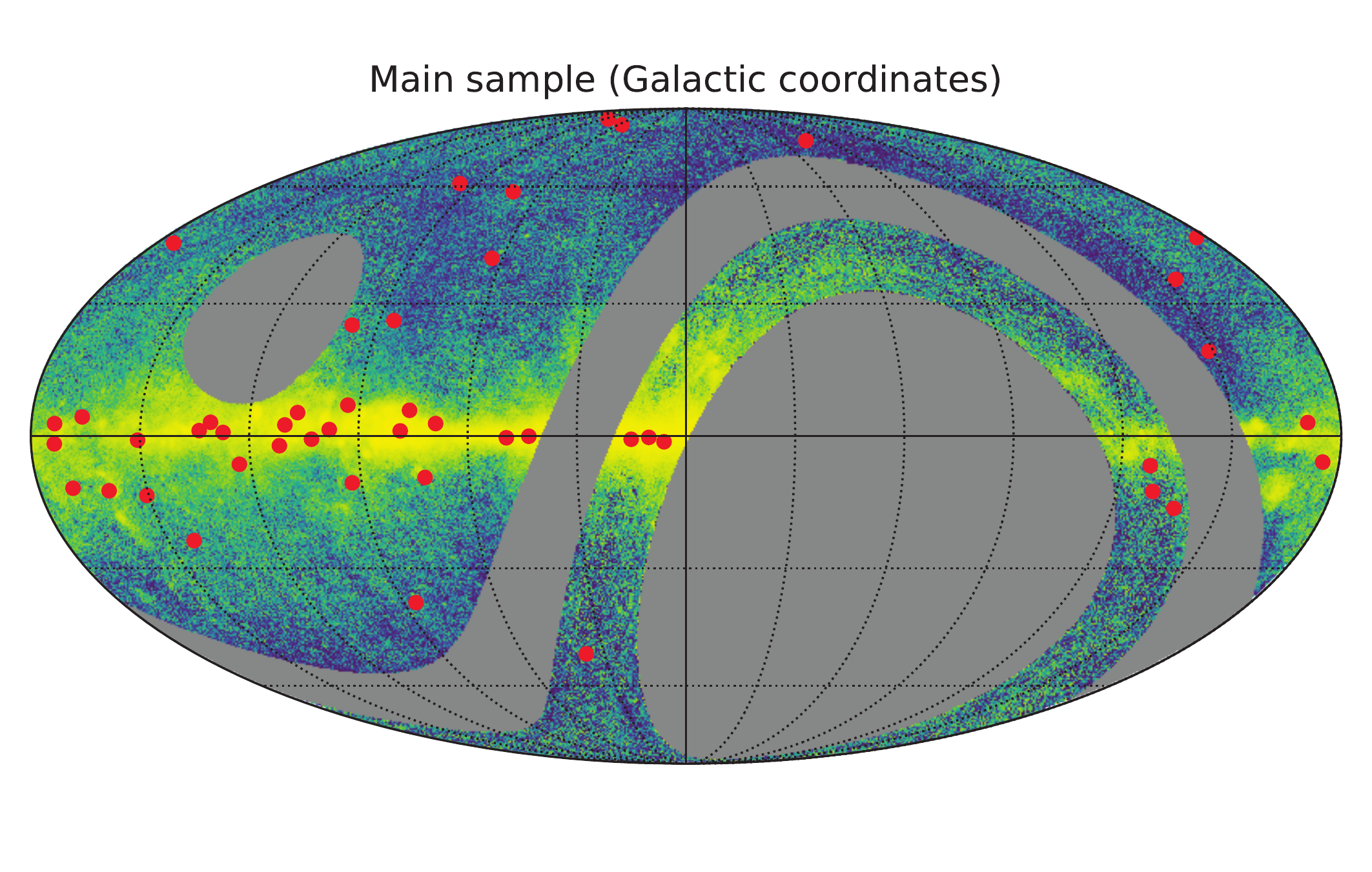}
     \end{subfigure}
     \hfill
     \begin{subfigure}[b]{\columnwidth}
         \centering
         \includegraphics[width=\columnwidth]{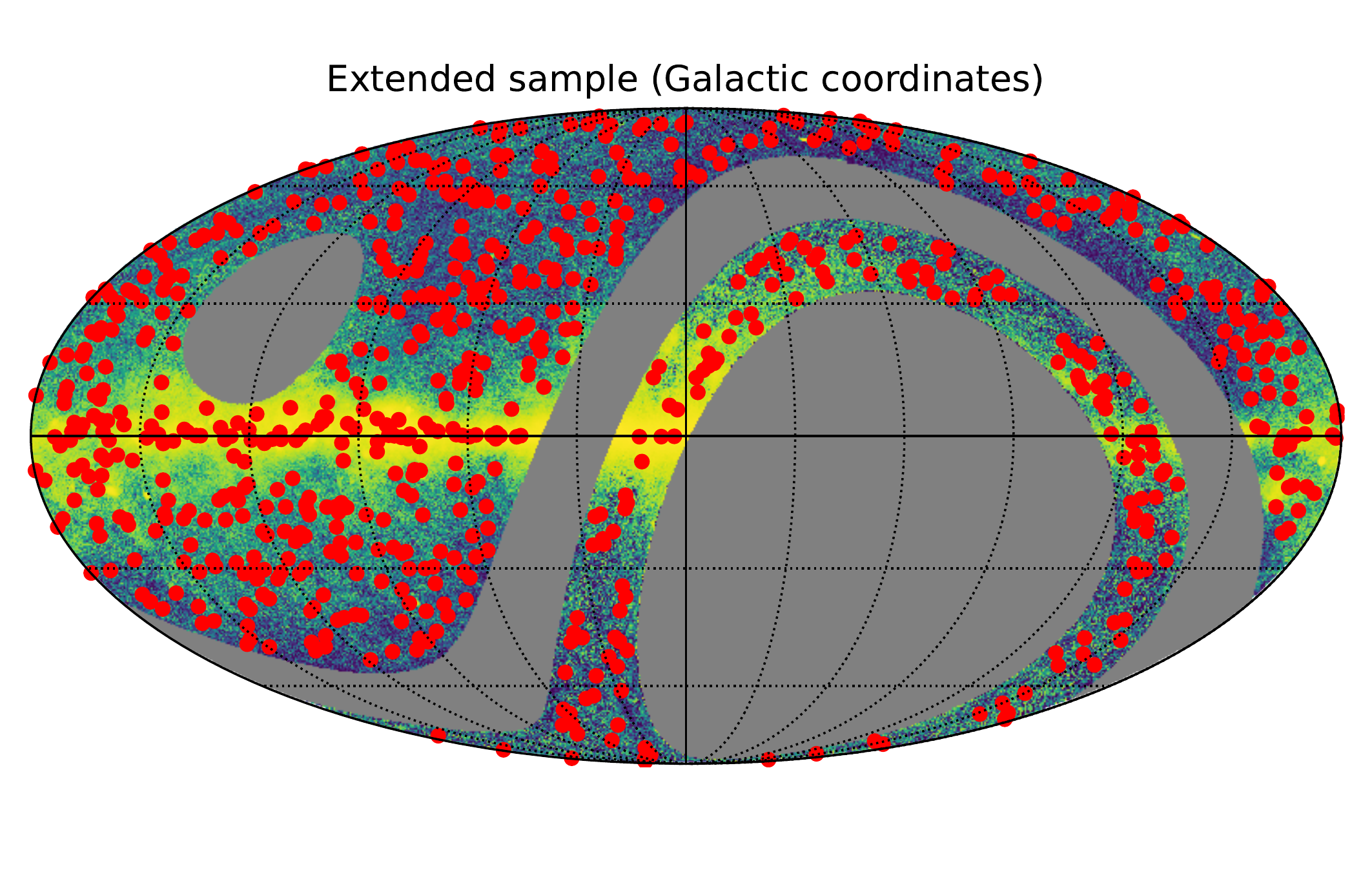}
     \end{subfigure}
     \hfill
     \setcounter{figure}{1} 
     \begin{subfigure}[b]{\columnwidth}
         \centering
         \includegraphics[width=\columnwidth]{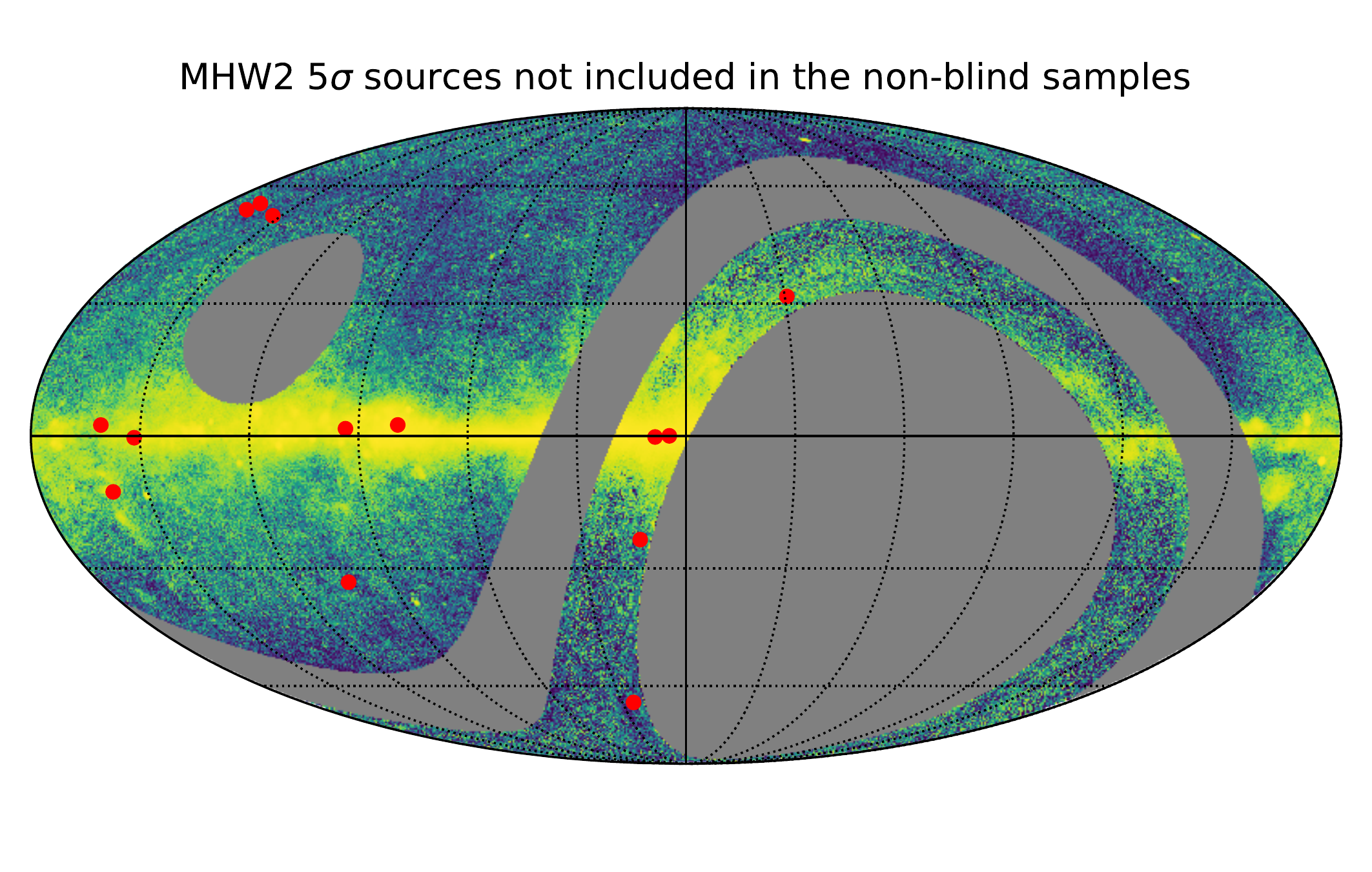}
     \end{subfigure}
    \caption{Positions in the sky, in Galactic coordinates, of 
our point source samples, superimposed on the QUIJOTE Wide Survey 
13 GHz sky. 
The grey area corresponds to our analysis mask, 
which is defined by the unobserved sky in the southern hemisphere, 
the region around declination zero degrees excluded owing to radio 
contamination from geostationary satellites, and the region around 
the north celestial pole with declinations above 70 degrees.
(sat+NCP  mask, see details in Rubi\~no-Mart\'\i n \textit{et al}. 2023). 
\textit{Top panel}: 
positions of the 47 sources in our main \textit{Planck} sample. 
\textit{Middle panel}: positions on the sky, in Galactic coordinates, 
of the 725 sources in our extended \textit{Planck} sample. 
\textit{Lower panel}: positions of the $14$ MHW2 $5\sigma$ targets 
not present neither in our main nor our 
extended samples.}  \label{fig:figs_one_to_three}
\end{figure}

We also study a non-blind extended sample (ES) that includes 
the positions of the remaining PCCS2 sources\footnote{That is, 
the PCCS2 sources with no detection in polarization or with 
a detection in polarization  with statistical confidence 
$<99.99\%$.} detected by \emph{Planck} at 30\,GHz \citep{PCCS2}
that lie in the area covered by the analysis mask proposed 
for the QUIJOTE MFI Wide Survey, and that are not in the main 
sample described above. We have used a search radius of one 
degree---slightly greater than  the QUIJOTE FWHM at 11\,GHz---to 
find and clean targets that may overlap at the QUIJOTE angular resolution. 
When a pair or group of possible repeated objects is found 
within this search radius, we keep only the one with the highest 
signal-to-noise ratio.
After this cleaning procedure, we keep 725 PCCS2 targets not 
already included in our main sample. From these 725 sources, 
145 ($20$ per cent)
 have Galactic latitude $|b|\leq 10^\circ$. We do not expect to 
provide  high-significance polarization measurements for most of 
this sample; however, the total intensity measurements for these 
sources should be useful for filling a gap  in the SEDs of many 
already known bright radio sources between low-frequency surveys 
such as the Parkes-MIT-NRAO Survey at 4.85\,GHz \citep[PMN,][]{PMN}, 
the Green Bank 6-cm  radio source catalogue, also at 4.85\,GHz 
\citep[GB6,][]{GB6}, CRATES at 8.4\,GHz \citep{CRATES},
 the Owens Valley Radio
Observatory (OVRO) blazar monitoring data at 15 GHz \citep{OVRO}, 
or the Arcminute Microkelvin Imager (AMI) Large Array, also at 15 GHz \citep{AMI},
 and higher frequency surveys such as the Australia Telescope 
20 GHz Survey \citep[AT20G,][]{AT20G}, and the WMAP  \citep{NEWPS,WMAP9yr} 
and \textit{Planck}  \citep{ERCSC,PCCS1,PCCS2,PCNT} point source 
catalogues. The middle panel of Figure~\ref{fig:figs_one_to_three} 
shows the positions on the sky, in Galactic coordinates, of the 
725 sources in the ES.

\subsection{Blind search} \label{sec:blind}

For the blind search, sources were detected at each frequency 
and horn independently, using improved versions of the detection 
pipelines used to create the first and second \emph{Planck} 
catalogues of compact sources \citep[PCCS and PCCS2,][]{PCCS1,PCCS2}. 
These pipelines are based on the Mexican Hat Wavelet 2 algorithm 
\citep[MHW2,][]{gnuevo06,caniego06}, which was also used to 
construct non-blind catalogues of sources from WMAP data 
\citep{NEWPS,NEWPS_study,NEWPS2}. MHW2 is a cleaning and denoising 
algorithm used to convolve the maps, preserving the amplitude of 
the sources while greatly reducing the large-scale structures 
visible at these frequencies (e.g.\ diffuse Galactic emission) 
and small-scale fluctuations (e.g.\ instrumental noise) in the 
vicinity of the sources.  
The preservation of the amplitude of the sources after filtering 
with  MHW2 allows us to use  MHW2 not only as a detector, but as 
an unbiased photometric estimator of the flux density of the 
sources. In the following, we shall refer to this photometric 
estimator as \textit{MHW2 photometry}.

The algorithm
projects the QUIJOTE full-sky temperature 
maps onto square patches where the filtering and detection is 
performed. The sizes of the patches ($14.658 \times 14.658$ deg$^2$), 
and the overlap between patches has been chosen in such a way 
that the full sky is effectively covered. Sources above a fixed 
signal-to-noise ratio (SNR) threshold are selected and their 
positions are translated from patch to spherical sky coordinates. 
Because the patches overlap, multiple detections of the same 
object can occur; these must be found and removed, keeping the 
detection with the highest SNR for inclusion in the catalogue. 
Because the MFI instrument observes the sky at 17 and 19\,GHz 
through two different horns,\footnote{The same can be said about 
11 and 13\,GHz, but the data from MFI horn 1 has not been used for 
this paper.} if a source is detected with both horns at those frequencies
the MHW2 catalogues 
 only keeps the detection of the horn with the higher SNR. 

The number of sources with {SNR$\geq 4$} detected by the MHW2 
are 178, 179, 145, and 143 at 11, 13, 17, and 19\,GHz respectively. 
For SNR$\geq 5$ the numbers drop to 118,
112,
74, and
50 sources at 11, 13, 17, and 19 GHz respectively. From those 
$5\sigma$ objects, 6, 4, 6, and 5 targets at 11, 13, 17, and 
19\,GHz respectively are not present in either the main sample or 
in the extended non-blind sample.\footnote{That is, these sources 
are not matched, within a one degree search radius, to any source 
in the main or extended non-blind samples.} Some of these 
detections correspond to the same source appearing above the 
$5\sigma$ level at different frequencies. In total, these 6+4+6+5 
detections correspond to 14 distinct positions on the sky.
The lower panel of  Figure~\ref{fig:figs_one_to_three} indicates 
the positions on the sky of those 14 blind $5 \sigma$ MHW2 
targets.\footnote{Note that a $5\sigma$ detection with MHW2 does 
not necessarily translate to a $5\sigma$ source in our catalogue.
This is because
MHW2  and the detection and estimation method used in this paper, 
as described in Section~\ref{sec:matched_filter}, use different 
algorithms to estimate photometric uncertainties. The algorithm 
used in this paper is more conservative and generally leads to 
greater uncertainties for the flux density and therefore smaller 
signal-to-noise ratios.} We keep their 14  unique coordinates as 
additional targets whose polarimetric properties will be studied 
with the technique described in Section~\ref{sec:FF} and that will 
be added to our extended sample of sources.
These 14 sources may be targets not considered in the PCCS2 owing
to the selection criteria followed by the \emph{Planck} collaboration.

In total, we are left with 47 sources from the original main 
sample,  725 sources from the extended sample and   14 targets 
from the blind MHW2 sample, leading to 786 targets to be studied 
in this paper. Most of these
targets have low signal-to-noise ratios in the MFI maps.
Table~\ref{tab:number_per_SNR} summarizes the number of sources 
detected in temperature (after filtering as described in 
Section~\ref{sec:matched_filter}) in our full sample (non-blind 
plus blind sources) above the $4\sigma$ and
$5\sigma$ levels in the whole sky outside the $|b|\leq 20^\circ$  
Galactic band.

\begin{table}
\begin{centering}
    \begin{tabular}{ccccc}
    \hline
    & 11 GHz & 13 GHz & 17 GHz & 19 GHz \\
    \hline
    N($\geq 4 \sigma$) &
149 (83) &
142 (81) &
81 (22) &
59 (12) \\
N($\geq 5 \sigma$) &
88 (42) &
85 (38)&
53 (12)&
36 (8) \\
\hline
    \end{tabular}
    \caption{Number of sources in our catalogue (blind + non-blind) detected in temperature above a given sigma level threshold
    in each of the MFI channels. Numbers between parenthesis indicate the number of
    sources above the same threshold that have Galactic latitude $b\geq 20^\circ$.}
    \label{tab:number_per_SNR}
    \end{centering}
\end{table}

\section{METHOD} \label{sec:FF}

Polarization of light is conveniently described in terms of the
Stokes parameters $\mathrm{I}$, $\mathrm{Q}$, $\mathrm{U}$ and 
$\mathrm{V}$ (see~\citealp{kamion} for a review on CMB polarization). 
The parameter $\mathrm{I}$ is the  total intensity of the radiation, 
whereas $\mathrm{Q}$ and $\mathrm{U}$ are the linear
polarization parameters, and $\mathrm{V}$ indicates the circular 
polarization,\footnote{For CMB photons, $\mathrm{V}$ is expected 
to be zero since Thomson scattering does not induce circular 
polarization. For this reason, throughout this paper we will consider V = 0.}
Whereas $\mathrm{Q}$ and $\mathrm{U}$ depend on the orientation 
of the reference frame. The
total polarization, defined as
\begin{equation} \label{eq:P}
\mathrm{P} \equiv \sqrt{\mathrm{Q}^2+\mathrm{U}^2},
\end{equation}
\noindent is invariant with respect to the relative orientation of
the receivers and the direction of the incoming signal, and
therefore has a clear physical meaning. 
Although. properly speaking. $\mathrm{Q}$ and $\mathrm{U}$ are components of a
$2\times2$ symmetric trace-free tensor, in practice the quantity 
$\mathrm{P}$ can  be treated
as the modulus of a vector.  \cite{FFpaper} studied the problem 
of the detection/estimation of a physical quantity that behaves 
as the modulus of a vector and is associated with a compact 
source embedded in stochastic noise. In particular, \cite{FFpaper}  
developed two techniques, one based on the Neyman-Pearson 
lemma \citep[see, for example,][]{herr10}---the Neyman-Pearson 
filter (NPF)---and another based on pre-filtering before fusion, 
the Filtered Fusion (FF), to deal with the problem of detection 
of the source and estimation of the polarization given two images 
corresponding to the $\mathrm{Q}$ and $\mathrm{U}$ components of polarization.
When the source is embedded in white Gaussian noise, the NPF and 
the FF are both easy to implement and perform similarly well 
\citep{FFpaper}, but for non-white noises the FF technique is 
much easier to implement---as it basically consists of the 
application of two independent matched filters on the $\mathrm{Q}$ 
and $\mathrm{U}$ images---so for this paper we have opted to use 
the Filtered Fusion technique. All the results presented in the 
following sections have been obtained with a Python version of the 
\texttt{IFCAPOL} code, which implements the FF technique and is 
publicly available through the RADIOFOREGROUNDS
portal.\footnote{\url{http://www.radioforegrounds.eu/pages/software/pointsourcedetection/ifcapol.php}}
The FF method and  \texttt{IFCAPOL} have already been used 
in the study of the polarization of WMAP \citep{WMAP_pol} 
and \emph{Planck} \citep{PCCS2} sources.

\subsection{Data model}

Let us consider a pair of images $d_{\mathrm{Q}}$ and 
$d_{\mathrm{U}}$  containing a point source plus some amount
 of noise, where by `noise' we mean any other physical or 
instrumental component apart from the point source itself 
(i.e.\ instrumental noise, Galactic and extragalactic foregrounds, 
CMB, etc.). The images could be all-sky spherical maps or local 
flat patches projected around a given celestial coordinate; in 
our experience, a local analysis helps to capture and to deal
with the non-stationarity of Galactic emission, so we therefore 
prefer to work with flat images. Each image has been obtained 
through an instrument with an angular beam response 
$\tau_y(\mathbf{x})$, where
$\mathbf{x})$ is the position on the sky
and
$y$ can take the values $\mathrm{I}$, $\mathrm{Q}$, $\mathrm{U}$ 
or, later on, $\mathrm{P}$ (note that the beam response does 
not need to be the same for the $\mathrm{I}$, $\mathrm{Q}$ and 
$\mathrm{U}$ images). Let us normalize the beam response so 
that $\tau_y(0) \equiv 1$, then for both images  we may write 
\begin{equation} \label{eq:datamodel}
  d_y(\mathbf{x}) = A_y\tau_y (\mathbf{x}) + n_y(\mathbf{x}),
\end{equation}
\noindent $A_y$ being the amplitude of the compact source  
$y \in (Q,U)$ and
 $n_y(\mathbf{x})$ the
corresponding noise in both components. Note that, in contrast 
to the situation for total intensity, where the amplitude is 
always positive, $A_y$ can have either sign depending on the 
polarization angle of the source. 

\subsection{Optimal filtering} \label{sec:matched_filter}

If the noises $n_y$ are Gaussian (but not necessarily white) 
the optimal estimator for $A_y$ is the matched filter 
\citep{kay,herr10}, which in Fourier space has the straightforward 
expression
\begin{equation} \label{eq:MF}
    \psi_y\left(\mathbf{k}\right) = \alpha_y \frac{\tau_y\left(\mathbf{k}\right)}{P_y(k)},
\end{equation}
\noindent
where $\mathbf{k}$ is the Fourier wave vector, 
$k\equiv |\mathbf{k}|$, $P_y(k)$ is the power spectrum of 
the noise $n_y(\mathbf{x})$ and $\alpha_y$ is an appropriate
 normalization. If we wish the filtered image to be an unbiased 
estimator of $A_y$ at the position of the source, we need
\begin{equation} \label{eq:norm}
    \alpha_y \equiv \left[ \int  d\mathbf{k} \,  \frac{\tau_y^2 \left(\mathbf{k}\right)}{P_y(k)} \right]^{-1}.
\end{equation}
\noindent
Equations (\ref{eq:MF}) and (\ref{eq:norm}) give the 
 most frequently used version of the matched filter in CMB astronomy.

If $\tilde{d}_{\mathrm{Q}}(\mathbf{x})$ and 
$\tilde{d}_{\mathrm{U}}(\mathbf{x})$ are the filtered images for 
the $\mathrm{Q}$ and $\mathrm{U}$ Stokes parameters and we have 
a source located at $\mathbf{x}=0$, then $\tilde{d}_{\mathrm{Q}}(0)$ 
and  $\tilde{d}_{\mathrm{U}}(0)$ are the optimal linear estimators 
of the amplitudes $A_{\mathrm{Q}}$ and $A_{\mathrm{U}}$, 
`optimal' meaning 
here that
\begin{itemize}
    \item $\tilde{d}_y(0)$ is an unbiased estimator of $A_y$; that 
is, $\langle \tilde{d}_y(0) \rangle = A_y$. The operator 
$\langle \cdot \rangle$ indicates an ensemble average.
    \item The ensemble variance  $\sigma^2_{\tilde{d}}$ is minimum.
\end{itemize}

The first of these two properties means that the matched 
filter can be used as an unbiased photometric estimator of 
the flux density of the sources, both in temperature and in 
the Q and U Stokes parameters. 
The r.m.s.\ of the filtered image in an annulus around the 
position of the sources can be used as an estimator of the  
uncertainty of the matched filter photometry.
Particular details about the size of the images and the annulus 
for this data set can be found in Section~\ref{sec:implementation}.
Unless stated otherwise, all the I, Q and U values, and their 
corresponding uncertainties provided in this paper are 
obtained using matched filter photometry (\textit{MF photometry}).

\subsection{Estimation of $\mathrm{P}$} \label{sec:Pestimation}

Once we have estimated the $\mathrm{Q}$ and $\mathrm{U}$ 
flux densities by means of the matched filters, we can construct 
a fusion-filtered map of $\mathrm{P}$ as
\begin{equation}
    \tilde{d}_{\mathrm{P}} (\mathbf{x})  = \sqrt{ \tilde{d}^2_{\mathrm{Q}} (\mathbf{x})
    + \tilde{d}^2_{\mathrm{U}}(\mathbf{x})}.
\end{equation}
For economy, let us change  our notation a little so that
\begin{equation}
\tilde{\mathrm{Q}}  \equiv  \tilde{d}_{\mathrm{Q}} (0) 
\end{equation}
\noindent
and
\begin{equation}
\tilde{\mathrm{U}}  \equiv  \tilde{d}_{\mathrm{U}} (0). 
\end{equation}
Then, over an ensemble of realizations, $\langle 
\tilde{\mathrm{Q}} \rangle = A_{\mathrm{Q}}$ and $\langle 
\tilde{\mathrm{U}} \rangle = A_{\mathrm{U}}$. It is 
straightforward to get the following estimator of $\mathrm{P}$:
\begin{equation} \label{eq:P_est}
    \tilde{\mathrm{P}} \equiv \sqrt{\tilde{\mathrm{Q}}^2 +  \tilde{\mathrm{U}}^2 }.
\end{equation}
\noindent
Unfortunately, this is a biased estimator of the amplitude 
of the polarization $A_{\mathrm{P}}$ of a source \citep[see 
for example][]{Montier}. \cite{FFpaper} studied the bias of 
the FF estimator (\ref{eq:P_est}), showing that it is easy 
to control for detectable sources. For  uncorrelated noises 
and assuming for simplicity that both images have the same 
r.m.s. noise level (for the QUIJOTE MFI Wide Survey maps this 
is a reasonable assumption), the relative bias can be shown 
to behave as
\begin{equation}
    \frac{\tilde{\mathrm{P}}-A_{\mathrm{P}}}{A_{\mathrm{P}}} \simeq \frac{1}{\mathrm{SNR}_{\mathrm{Q}}^2+\mathrm{SNR}_{\mathrm{U}}^2},
\end{equation}
\noindent
where SNR$_y$ is the signal-to-noise ratio of the source for 
the Stokes parameter $y$ after filtering. This means that if 
the source is detectable at the $3\sigma$ level in at least 
$\tilde{\mathrm{Q}}$ or $\tilde{\mathrm{U}}$ the relative 
bias should be less than or equal to $\sim$10\%, and if at least 
one of them  is at the  $5\sigma$ level the relative bias 
would be $\leq 4\%$. In practice, the bias will be negligible 
for the brightest sources.
In any case, the noise bias can be removed by subtracting the corresponding noise
contributions $\sigma_{\tilde{\mathrm{Q}}}^2$ and 
$\sigma_{\tilde{\mathrm{U}}}^2$ from 
the filtered $\tilde{\mathrm{Q}}^2$ and $\tilde{\mathrm{U}}^2$ 
images . As noted by \cite{WMAP_pol}, 
this correction turns out to be negligible in most cases. We 
shall provide the debiased polarized flux density as
\begin{equation} \label{eq:P_debiased}
    \tilde{\mathrm{P}}_{\mathrm{debiased}} = 
\sqrt{\tilde{\mathrm{P}}^2-\sigma_{\tilde{\mathrm{P}}}^2},
\end{equation}
\noindent
where $\sigma_{\tilde{\mathrm{P}}}$ is the error in 
$\tilde{\mathrm{P}}$  and is calculated by propagating the 
errors in $\tilde{\mathrm{Q}}$ and $\tilde{\mathrm{U}}$. 
The standard deviations $\sigma_{ \tilde{\mathrm{Q}} }$ and 
$\sigma_{ \tilde{\mathrm{U}} }$ are calculated as the local 
r.m.s.\ in an annulus around the source in the matched filtered 
$\tilde{\mathrm{Q}}$ and $\tilde{\mathrm{U}}$ maps. Then, 
under the assumption of no correlation:
\begin{equation}
    \sigma_{\tilde{\mathrm{P}}} = \sqrt{\frac{\tilde{\mathrm{Q}}^2 \, \sigma_{\tilde{\mathrm{U}}}^2 + \tilde{\mathrm{U}}^2 \, \sigma_{\tilde{\mathrm{Q}}}^2}{\tilde{\mathrm{Q}}^2+\tilde{\mathrm{U}}^2}}.
\end{equation}
\noindent
Even if the noises $n_{\mathrm{Q}}(\mathbf{x})$ and 
$n_{\mathrm{U}}(\mathbf{x})$ are Gaussian-distributed, the 
noise residual after the filtered fusion is not Gaussian. 
Rather than working with the usual signal-to-noise ratio, it 
is more appropriate to report the statistical significance of 
the detection of $\mathrm{P}$; that is,
one minus the probability that a given estimated signal is  
the product of random noise fluctuations. We compute the 
statistical significance by obtaining histograms of the 
$\tilde{d}_{\mathrm{P}}$ values,  as described in \cite{WMAP_pol}.

\subsection{Estimation of the polarization angle}

From the filtered $\mathrm{Q}$
 and $\mathrm{U}$ images we can estimate the angle of 
polarization of a source located at $\mathbf{x}=0$ as
 \begin{equation} \label{eq:polangle}
     \tilde{\phi} = \frac{1}{2} \arctan \left[-\frac{ \tilde{\mathrm{U}}}{\tilde{\mathrm{Q}}} \right].
 \end{equation}
 \noindent
The QUIJOTE MFI maps use the COSMO polarization convention. 
The minus sign in equation (\ref{eq:polangle}) is inserted 
to obtain the $\tilde{\phi}$ angle in the IAU convention, 
so that we can compare it with other experiments.  
Assuming no correlation between the errors in $\tilde{\mathrm{Q}}$ 
and $\tilde{\mathrm{U}}$,
\begin{equation}
    \sigma_{\tilde{\phi}} =
    \frac{1}{2\left(\tilde{\mathrm{Q}}^2 +
 \tilde{\mathrm{U}}^2 \right)} \sqrt{ \tilde{\mathrm{Q}}^2 \, \sigma_{\tilde{\mathrm{U}}}^2 + \tilde{\mathrm{U}}^2 \, \sigma_{\tilde{\mathrm{Q}}}^2}.
\end{equation}

\subsection{Implementation} \label{sec:implementation}

\begin{figure*}
	\includegraphics[width=\textwidth]{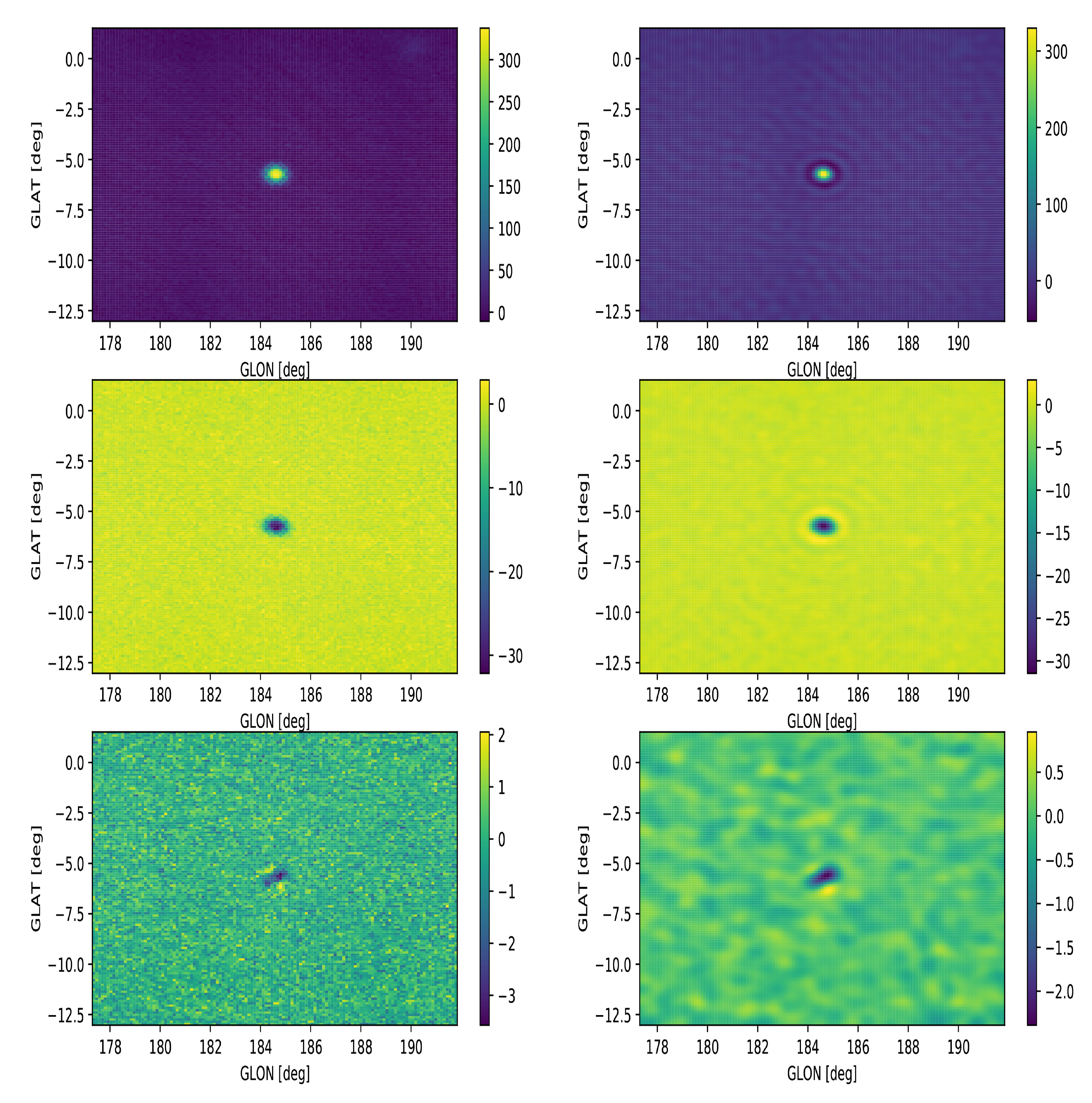}
    \caption{Intensity and polarization maps at 19 GHz centred 
on the position of the Crab nebula. The upper left- and right-hand
 panels show, respectively, the maps of $\mathrm{I}$  and 
$\tilde{\mathrm{I}}$ before and after filtering with the matched 
filter, while the middle panels show the analogous $\mathrm{Q}$ and 
$\tilde{\mathrm{Q}}$ maps. The lower panels
   show, for the same source, the maps of $\mathrm{U}$ and $\tilde{\mathrm{U}}$.}
    \label{fig:Crab}
\end{figure*}

The \texttt{IFCAPOL} version used for this work produces flat 
patches of $128 \times 128$  pixels using the Gnomonic projection 
integrated in the Python \texttt{healpy} package \citep{healpy} 
based on the \texttt{HEALPix} \citep[Hierarchical Equal Area 
isoLatitude Pixelation of a sphere,][]{healpix}.\footnote{\url{http://healpix.sourceforge.net}}
For \texttt{nside} = 512 \texttt{HEALPix} maps,
each pixel has an angular resolution of 6.87 arcmin; that 
is, each patch covers an area of $14.658 \times 14.658$ deg$^2$. 
The FF uses an isotropic, non-Gaussian model of the beam response 
$\tau(\mathbf{x})$ obtained from the window functions described 
in \cite{MFIpipeline}. The mean and r.m.s.\ values used for SNR 
calculations and $\tilde{\mathrm{P}}$ debiasing are calculated in 
rings with inner radius 2.5 times the nominal FWHM of the beam and 
with an outer radius equal to 6 degrees. The significance of 
$\tilde{\mathrm{P}}$ is calculated over all the pixels of the image, 
except for a 5-pixel band in the borders of the image (to avoid 
filtering border effects) and a mask that covers  the $2.5\%$ 
fraction of brightest pixels of the image outside the area occupied 
by the source
(in a similar way to \citealp{WMAP_pol}, but with a more conservative 
criterion for masking bright pixels)
to prevent contaminated pixels from being used in the calculation of the 
background distribution. This means that in the best case the 
reliability of a detection cannot be ascertained 
to a level higher than $99.993\%$.
Figure~\ref{fig:Crab}  shows an example of a patch located around 
the Crab nebula in I, Q and U before (left panels) and after (right 
panels) filtering. Ringing effects around bright sources, such as 
those observed in the filtered Q map, are a side effect of 
filtering. For this reason, an annulus around the sources is used, 
as mentioned at the beginning of this section. \rtgs{The lobes 
in the U map is an instrumental effect associated with 
intensity-to-polarization leakage due to the difference between 
the two co-polar beams, as explained in section 9.3 of \cite{mfiwidesurvey}.
 These are visible in this case because most of Crab polarized
 emission is contained in the Q direction in Galactic coordinates}.

The polarization angle is calculated using Equation (\ref{eq:polangle}).
Since the noise can easily change the sign of $\mathrm{Q}$ or 
$\mathrm{U}$, and hence $\hat{\phi}$, in cases of very low SNR sources,
 this implementation \texttt{IFCAPOL} tries automatically to 
homogenize the sign of $\tilde{\phi}$.
 It does this via the following
 ad hoc rule: for a given source, the sign 
of 
its polarization angle for all frequencies in the final catalogue 
is forced to be equal to the 
sign of the median value 
of the polarization angles at 11, 13, 17 and 19 GHz as calculated 
using Equation (\ref{eq:polangle}).
This is a brute force solution; however, it is necessary since only 
four sources in the main sample have SNR $\geq 1.0$ in both $\mathrm{Q}$ 
and $\mathrm{U}$ at all frequencies. All these four sources have 
Galactic latitude $|b|<10^\circ$. These sources are listed in 
Table~\ref{tab:high_SNR}.

\begin{table}
\begin{tabular}{llrr}
\hline
PCCS2 ID & Other ID & RA [deg] & DEC [deg] \\
\hline
\hline
PCCS2 030 G184.54-05.78 & Crab & 83.63 & 22.01 \\
PCCS2 030 G130.72+03.08 & 3C058 & 31.40 & 64.82 \\
PCCS2 030 G189.07+03.02 & IC443 & 94.30 & 22.55 \\
PCCS2 030 G068.83+02.80  & CTB 80 & 298.31 &  32.92 \\
\hline
\end{tabular}
\caption{List of highly polarized  sources in the main sample. These sources have SNR $\geq 1.0 \sigma$ in $\mathrm{Q}$ and $\mathrm{U}$ at all MFI frequencies.}
\label{tab:high_SNR}
\end{table}

\subsection{Spectral indexes}
\label{sec:cc}

\jarm{In order to obtain the estimates of the spectral indexes 
based on internal QUIJOTE MFI data only, it is important to include 
colour corrections in the analysis. Even though these are small 
corrections (of order of one  per cent) they might bias the spectral 
index estimate given the small frequency range considered (11 to 19\,GHz).}  
The MFI procedure to obtain colour corrections is described in 
section 8.2 of \cite{MFIpipeline} and implemented by the 
{\sc fastcc}\footnote{\url{https://github.com/mpeel/fastcc}} 
(faster computation of the colour correction for a given spectral 
index) code \citep{fastcc}. 
For point sources in this paper, we consider the spectral 
behaviour of bright radio sources to be well described by  a 
single power law \jarm{within the MFI frequency range}. We 
follow an iterative procedure according to this scheme:
\begin{enumerate}
\renewcommand{\labelenumi}{\roman{enumi}.}
    \item For each source, we take the four photometric points 
(I or P at 11, 13, 17 and 19 GHz) and their corresponding 
uncertainties as an initial estimate. 
    \item We then get an  estimate of the spectral index 
by a weighted fitting of the points to a power law.
    \item Using the estimated spectral index, we compute the 
colour correction factors for the four frequencies.
    \item We apply these corrections to the 
non-colour-corrected I or P points and repeat the process from 
step ii. We iterate until the resulting spectral index 
stabilizes to within a $10^{-6}$ tolerance. 
\end{enumerate}
\noindent
The process typically requires 5--7 iterations to achieve 
the requested tolerance \jarm{on the spectral index.  
MFI colour corrections are typically small. }
For the Stokes parameter I, the largest effect occurs at 
11 GHz, with corrections of up to $\sim$3\% for sources 
with spectral indices between $-1$ and $+1$. This correction 
remains below $\sim$0.5\% for 13\,GHz and below $\sim$1\% 
for 17 and 19\,GHz, for the same spectral index range. Similar 
values are obtained for Q, U and P. 

\jarm{As a consistency check, we have implemented an 
independent method to fit for the spectral index, using 
a Bayesian approach that includes the simultaneous 
evaluation of the colour correction factors within the 
computation of the posterior distribution for the spectral 
index. As in our default methodology, flux densities are 
fitted to a power law, $S(\nu;A_0,\alpha) =
 A_0 \left( \nu/\nu_0 \right)^{\alpha}$ (normalized 
at $\nu_0=$10\,GHz). This approach is applied only to 
the intensity signal in order to validate the previous 
results (spectral indexes and colour correction factors).}
\jarm{The posterior distribution for the spectral index is 
sampled using {\sc emcee},\footnote{\url{https://emcee.readthedocs.io/en/stable/}} 
\citep{2013PASP..125..306F} following the methodology explained 
in \cite{snrwidesurvey}.} 
\jarm{A uniform prior on $\alpha$ between $-$4 and 3 is adopted. 
For each source, the full posterior distribution function (PDF) 
is characterized with 32 chains and 10\,000 iteration steps.}
Then, $\alpha$ and $A_0$ are estimated from the 50th percentile 
of the marginalized PDFs, while their uncertainties are 
estimated from the 16th and 84th percentiles.
In addition, the full PDFs are a useful and complementary tool 
in the statistical description of our sources 
(see Sect.~\ref{sec:spectral_index}).
The results of this second method (MCMC sampler) are 
virtually identical to those  obtained with the iterative 
power-law fitting described above.
Small differences arise from the parameter estimation methods, 
since the latter estimates the  parameters as the best-fitting 
value (corresponding to the maximum of the PDF) while the 
former obtains them from the median. The mean difference 
between spectral indexes is $-0.07$ with a standard deviation 
of 0.08, where the biggest difference is $-0.47$.
However, we note that the MCMC analysis generally provides 
greater uncertainties, which on average are a factor 2.57 higher.

\chlc{Concerning} the impact of colour correction on the estimation 
of the spectral indices of the sources, we find that the average 
difference  between colour-corrected $\alpha$ and 
non-colour-corrected $\alpha$ is 0.06, which is smaller than 
the average $\alpha$ uncertainties of 0.39 \chlc{(using the 
MCMC results)}. Although this effect is small, it may have a 
significant impact on the extrapolation of flux densities to 
lower or higher frequencies.
The subsequent analysis will therefore be done using the 
colour-corrected flux densities.

\section{Description and validation of the data products} \label{sec:products}

\subsection{Format of the data products}

The QUIJOTE MFI Wide Survey point source catalogue is 
available from the RADIOFOREGROUNDS web 
site.\footnote{\url{http://www.radioforegrounds.eu/pages/data-products.php}} 
It comprises two catalogue FITS files, one 
for our main sample, containing 47 targets, and other 
for our extended sample, containing 739 targets. 
We summarize here the catalogue contents:
\begin{itemize}
\renewcommand\labelitemi{--}
    \item Source identification: a numerical identifier 
containing the string \texttt{`MFI-PSCm'} for the main sample 
and \texttt{`MFI-PSCe'} the extended sample.
    \item Position: GLON and GLAT contain the Galactic 
coordinates, and RA and DEC give the same position 
in equatorial coordinates (J2000).
    \item Stokes parameters: I, Q and U in Jy, and their 
associated uncertainties, for the four MFI frequencies 
(11, 13, 17 and 19 GHz). Values are not colour-corrected. 
Colour-corrections can be obtained using the public 
\texttt{fastcc} code (Peel et al., in preparation).
    \item Polarization: debiased P and its associated 
uncertainty in Jy for the four MFI frequencies 
(11, 13, 17 and 19 GHz).  Values are calculated from 
the raw (non-colour-corrected) Stokes parameters.
    \item Polarization fraction and its associated 
uncertainty for the four MFI frequencies, as calculated 
from debiased P and I.
    \item Polarization angle: $\phi$ and its associated 
uncertainty, in degrees, for the four MFI frequencies 
(11, 13, 17 and 19 GHz).   The polarization angles are 
defined as increasing anticlockwise (north through east) 
following the IAU convention; the position angle zero 
is the direction of the north Galactic pole.
    \item Statistical significance of the detection of 
the  polarized signal, for the four MFI frequencies.
    \item Spectral index in intensity: column 
\texttt{ALPHA\_I} gives the colour-corrected spectral 
index calculated as described in section~\ref{sec:cc}. 
Its associated error is given in column  \texttt{ALPHA\_I err}.
    \item Spectral index in polarization: column 
\texttt{ALPHA\_P} gives the colour-corrected spectral 
index calculated as described in section~\ref{sec:cc}. 
Its associated error is given in column  \texttt{ALPHA\_P err}.
    \item Flag:  source candidates
         with estimated $\mathrm{SN}R < 0$ in at least one  frequency 
        are flagged with the number 1. For these sources, 
the corresponding I column has been set to NaN. The 
corresponding uncertainty is kept as it may be used as 
an upper flux density limit estimate.  We do not provide 
spectral index estimates for these sources, as they are 
considered as only marginal detections.   There are 14 of 
these sources in the extended catalogue and zero in the 
main catalogue. The rest of sources are flagged with the value 0.
    \item Cross-identifications: where possible,  
we give the name of possible cross-identifications, 
within a $30^\prime$ search radius around the position of each
 MFI source candidate, to the following surveys of radio sources:
    \begin{itemize}
        \item \texttt{PCCS2 ID}: nearest matched source in the 
\emph{Planck} Second Catalogue of Compact Sources \citep{PCCS2}.
        \item \texttt{PNCT ID}: nearest matched source in the 
\emph{Planck} Catalogue of Non-Thermal Sources \citep{PCNT}.
        \item \texttt{Other IDs}: for those cases where it is
 possible, we also give the nearest match, within a  $30^\prime$ 
search radius, to the 3C \citep{3C} and the 
    Parkes-MIT-NRAO   \citep[PMN,][]{PMN} surveys of radio sources.
    \end{itemize}
\end{itemize}

\subsection{Internal consistency} \label{sec:consistency}

In order to check the self-consistency of \texttt{IFCAPOL}, 
as well as have at hand an additional means of testing the stability of 
QUIJOTE data and the map-making algorithm 
\citep[see][]{MFIcontrolsystem,status2016SPIE,mfiwidesurvey,MFIpipeline}, 
we have conducted a series of jackknife tests comparing 
the estimation of the main photometric and polarimetric 
quantities ($\mathrm{I}$, $\mathrm{P}$ and $\phi$, but 
also $\mathrm{Q}$ and $\mathrm{U}$ individually) on two 
different maps, which have been produced by splitting the 
MFI wide survey data into two halves. 
The maps that we use are the so-called `half-mission maps' 
and they are constructed 
\jarm{by separating the individual observations (6 h 
each) according to the calendar date in each observing
period (there are four in total, spread over 6 yr) and each 
telescope elevation (usually we have three or four observing
 elevations in each period). A more complete description of 
these maps is given in \cite{mfiwidesurvey}. By construction, 
these maps are designed to have almost identical sky signal 
and sky coverage, but  independent noise. Note that by design, 
the effective epoch of the two half-mission maps is almost identical. }
We use specific versions of these half-mission maps 
that have been generated with our map-making algorithm 
\citep[PICASSO,][]{destriper} using the common baseline 
reconstruction as for the full map \citep{mfiwidesurvey}. 
This guarantees a better rejection of $1/f$ noise in the 
two halves, which benefits the analyses that are presented 
in this section.

\begin{figure*}
	\includegraphics[width=\textwidth]{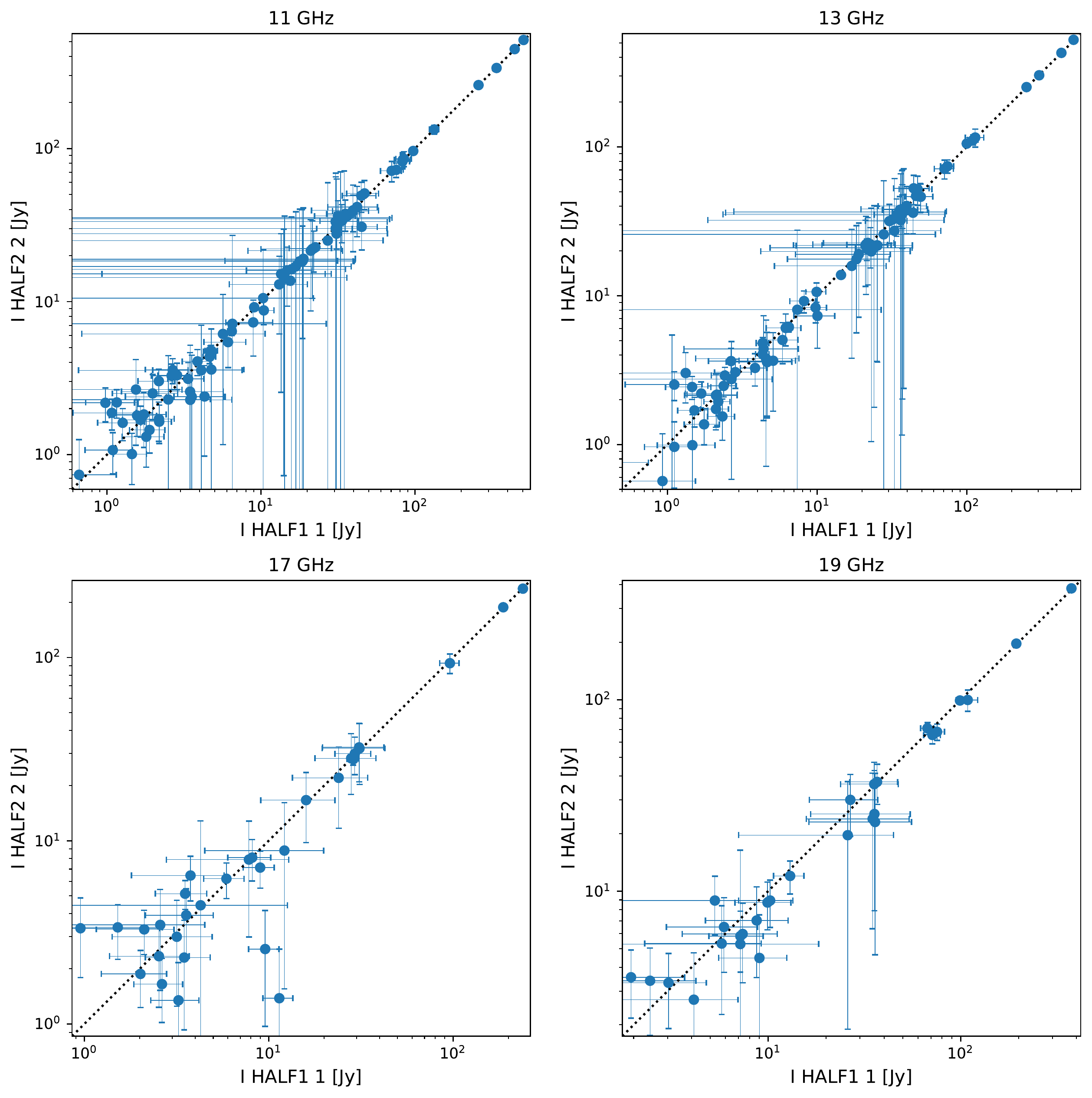}
    \caption{Comparison of $\mathrm{I}$ estimation in the two
 MFI Wide Survey half-surveys. The black dotted line indicates 
the equality $y=x$.}
    \label{fig:I_half_half}
\end{figure*}

\begin{figure*}
	\includegraphics[width=\textwidth]{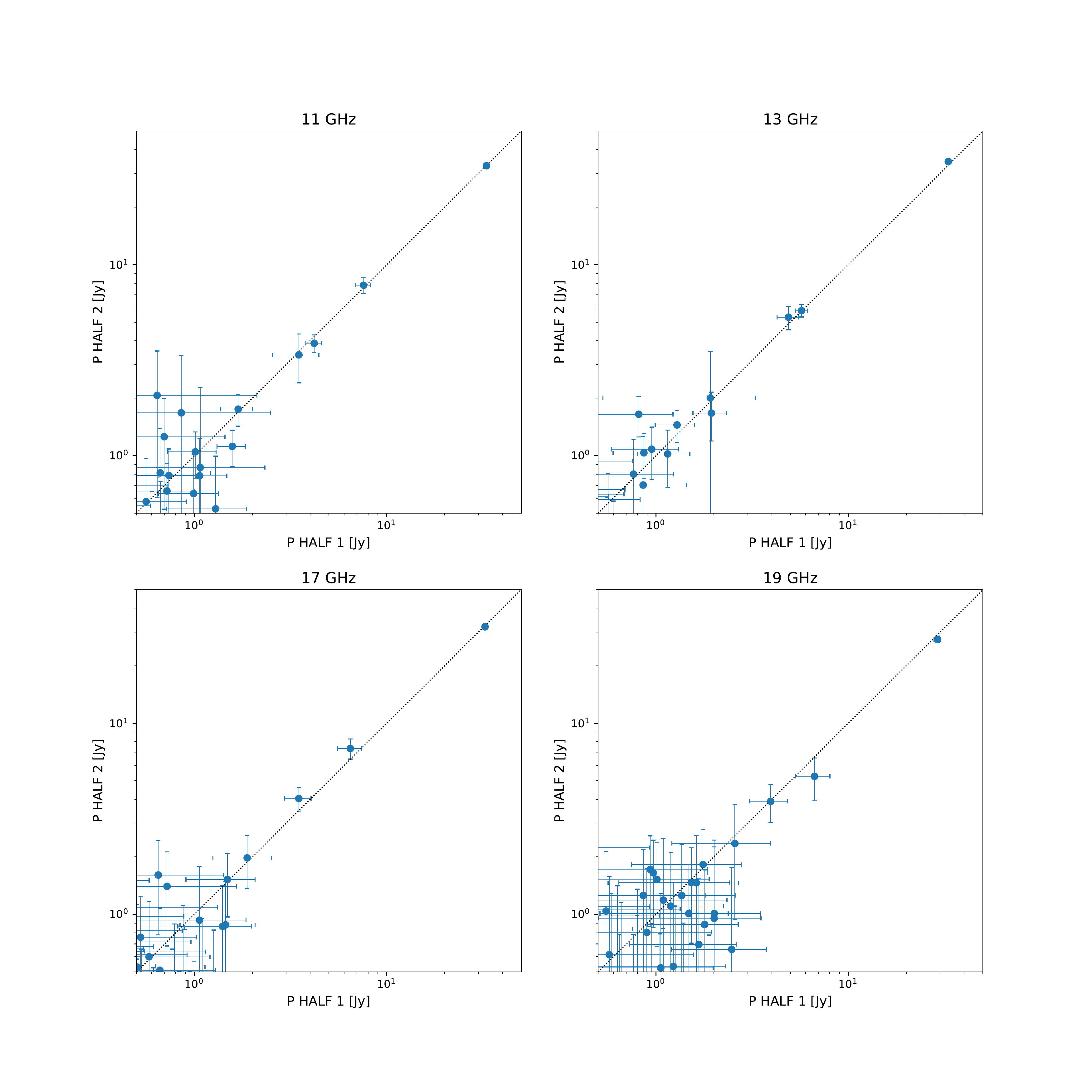}
    \caption{Comparison of $\mathrm{P}$ estimation in the two 
MFI Wide Survey half-surveys. The black dotted line indicates 
the equality $y=x$.}
    \label{fig:P_half_half}
\end{figure*}

\begin{figure*}
	\includegraphics[width=\textwidth]{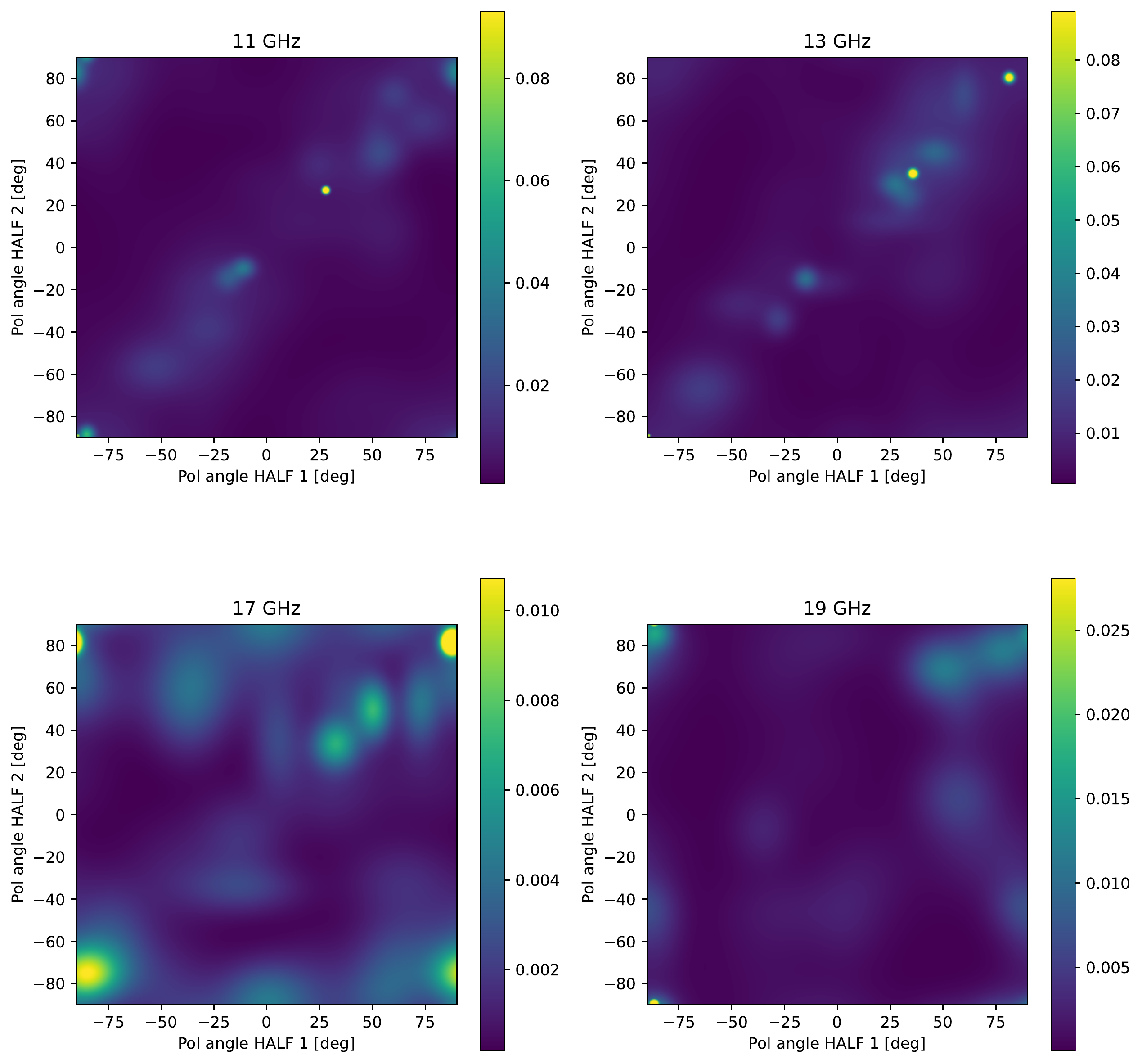}
    \caption{Comparison of the polarization angle estimation in 
the two MFI Wide Survey half-surveys, shown as a density plot.}
    \label{fig:angle_half_half}
\end{figure*}

Figures~\ref{fig:I_half_half},~\ref{fig:P_half_half} 
and~\ref{fig:angle_half_half} show how the two half-survey 
maps compare when we estimate $\mathrm{I}$, $\mathrm{P}$, and $\phi$
respectively. In order to
reduce spurious scatter from the use of different horns,
 only values that have been obtained using the same MFI horn
 are used for the plots. An additional cut requiring that 
the measurement of P has statistical significance $\geq 0.68$ 
in the full map has been applied for the comparison. The 
numbers of sources satisfying the above requirements, 
 and are therefore used for the fit, are 81, 75, 31 and 26 for 
11, 13, 17 and 19 GHz respectively.
  A summary of the results of a linear fit between the 
estimates obtained with the first and the second half-survey 
is presented in Table~\ref{tab:fit_half_half}. 
  The I and P estimations show good concordance between the 
subsets used for this test. The 17 and 19\,GHz channels have 
fewer samples and show a wider scatter
  than the lower frequencies, probably because steep-spectrum
 sources decrease quickly in flux density as the observation 
frequency increases. Both half-survey maps show similar sensitivities, 
as expected since each of them corresponds to the same number 
of samples. This is confirmed by the size of the error bars and 
by a quantitative analysis of these uncertainties, which shows 
no statistically significant discrepancies between both half-survey maps.

 \begin{table*}
\begin{tabular}{cccc} \hline
Freq [GHz] & I [Jy] & P [Jy] & $\phi$ [deg] \\
11 & $(0.996 \pm 0.008)x + (0.07 \pm 0.13)$ & $(0.99 \pm 0.02)x - (0.00 \pm 0.03)$ & $(1.00 \pm 0.01)x - (0.2 \pm 0.8)$ \\
13 & $(1.002 \pm 0.008)x + (0.08 \pm 0.14)$ & $(1.04 \pm 0.02)x - (0.02 \pm 0.04)$ & $(1.00 \pm 0.01)x + (0.0 \pm 0.8)$ \\
17 & $(0.996 \pm 0.014)x - (0.4 \pm 0.4)$ & $(0.89 \pm 0.15)x + (0.02 \pm 0.09)$ & $(1.03 \pm 0.05)x - (5.0 \pm 3.0)$ \\
19 & $(1.015 \pm 0.010)x - (0.4 \pm 0.9)$ & $(1.01 \pm 0.04)x - (0.09 \pm 0.14)$ & $(1.21 \pm 0.04)x + (16.0 \pm 3.0)$ \\
\hline \end{tabular} \caption{Fit coefficients between the first and second half-survey maps for the total intensity, polarization and polarization angle of the main sample catalogue.}  \label{tab:fit_half_half}
\end{table*}

 Regarding the polarization angle,  the estimated angles show 
large error bars and, even after the homogeneization process 
described in Section~\ref{sec:implementation}, a significant 
scatter. Error-bar diagrams such as those shown in 
Figures~\ref{fig:I_half_half} and~\ref{fig:P_half_half} become
 very cluttered by the error bars and difficult to read. We have
chosen instead to show a density diagram
 in  Figure~\ref{fig:angle_half_half}. 
 For this diagram, each point has been substituted by a normalized 
Gaussian ellipsoid with 
 semiaxis equal to the estimated polarization angle error. 
Well-determined polarization angles create dense knots in this 
diagram, whereas sources with a large uncertainty 
 form very spread out clouds. There is a clear linear trend at 
11, 13 and, to a lower extent, 17 GHz. There is a cluster of 
sources with switched angle in the upper left corner of the plot 
at 19 GHz.
 The reason for this is that
 when either $\mathrm{Q}$ or $\mathrm{U}$ is small, random 
noise fluctuations can change the sign or the semi-quadrant 
of $\tilde{\phi}$. This is particularly dangerous for sources 
with polarization angle near $\pm 90$ degrees.  If we artificially 
force the signs of angles whose absolute value is larger than 80 
 degrees to agree, the 19 GHz  fit becomes $\tilde{\phi}_{H2} = 
(0.95\pm 0.02)\tilde{\phi}_{H1} - (7.59 \pm 2.04)$. Here the 
H1 and H2 subscripts denote the first and second half-mission 
maps respectively. In any case, Figure~\ref{fig:angle_half_half}  
indicates that polarization angle estimates at 17 and 19 GHz 
are less reliable than those at 11 and 13 GHz.

\subsection{MF versus MHW2 photometry} \label{sec:MF_vs_MHW2}

As mentioned in Sections~\ref{sec:blind} and~\ref{sec:implementation}, 
both filters used for the non-blind (MF) and blind (MHW2) samples are,
from a statistical point of view, detectors and estimators of the 
flux density of the sources at one and the same time. We use the MF 
photometry for all the products that come with this paper, 
except for the number-count analysis to be described in 
Section~\ref{sec:number_counts}, where we have opted to keep the 
native MHW2 photometry of the blind sample for statistical 
consistency. It is still interesting to test whether the MHW2 
and the MF photometries are compatible. We have fitted the MF 
estimated intensity  versus the corresponding MHW2 photometry 
for all the matches between the blind and non-blind sources, 
taking into account the photometric uncertainties in both axes. 
Since all the detections are internal to the QUIJOTE MFI (we are 
not comparing with external catalogues), here we allow the 
matching radius to be roughly equal to the FWHM of the 
best-resolution MFI channel ($\sim$35 arcmin at 17 and 19 GHz).
Table~\ref{tab:MF_vs_MHW2} shows
the linear-fit parameters and their uncertainties for sources
 with  $\mathrm{SNR} \geq 5$. The agreement is remarkable.
 Since both photometries have been obtained with two totally 
independent codes and the MHW2 has been well tested in other 
experiments such as WMAP and \emph{Planck}, we take this 
result as an additional validation of our photometry.

\begin{table}
\begin{center}
\begin{tabular}{ccc}
\hline
Freq [GHz] &  $a$ &  $b$ [Jy] \\
\hline
  11 & $1.01 \pm  0.01$ & $0.01 \pm 0.07 $ \\
  13 & $1.00 \pm 0.01$ & $0.01 \pm 0.07 $ \\
  17 & $1.02 \pm  0.01$ & $-0.42 \pm 0.21 $\\
  19 & $1.01 \pm 0.01$ & $-0.40 \pm 0.43$ \\
\hline\end{tabular}
\caption{Parameters of the fit 
$\mathrm{I}_{\mathrm{MHW2}} = a \, \mathrm{I}_{\mathrm{MF}} + b$
between the MHW2 photometry and the MF photometry, 
for all the sources in the blind sample that are matched 
 within a 35 arc minutes search radius
to a $\mathrm{SNR} \geq 5$
source in the non-blind sample.} \label{tab:MF_vs_MHW2}
\end{center}
\end{table}

\subsection{Calibrators} \label{sec:calibrators}

The main amplitude calibrator of the QUIJOTE MFI Wide 
Survey is the Crab supernova remnant (Tau A), while the 
supernova remnant Cassiopeia A (Cas A) and the radio 
galaxy 3C405 (Cygnus A) are used for 
consistency tests. This calibration is based on
 beam-fitting photometry, performed at either the time-ordered 
data or at the map level---see details in \cite{mfiwidesurvey},
and a general description 
of the QUIJOTE MFI calibration in \cite{MFIpipeline}. 
Comparison of the reference flux densities of these sources 
at the MFI frequencies, given by the external models that
 were used for calibration, is a very useful validation test 
of our methodology, which relies on a different and 
independent kind of photometry.

\begin{figure*}
	\includegraphics[width=\textwidth]{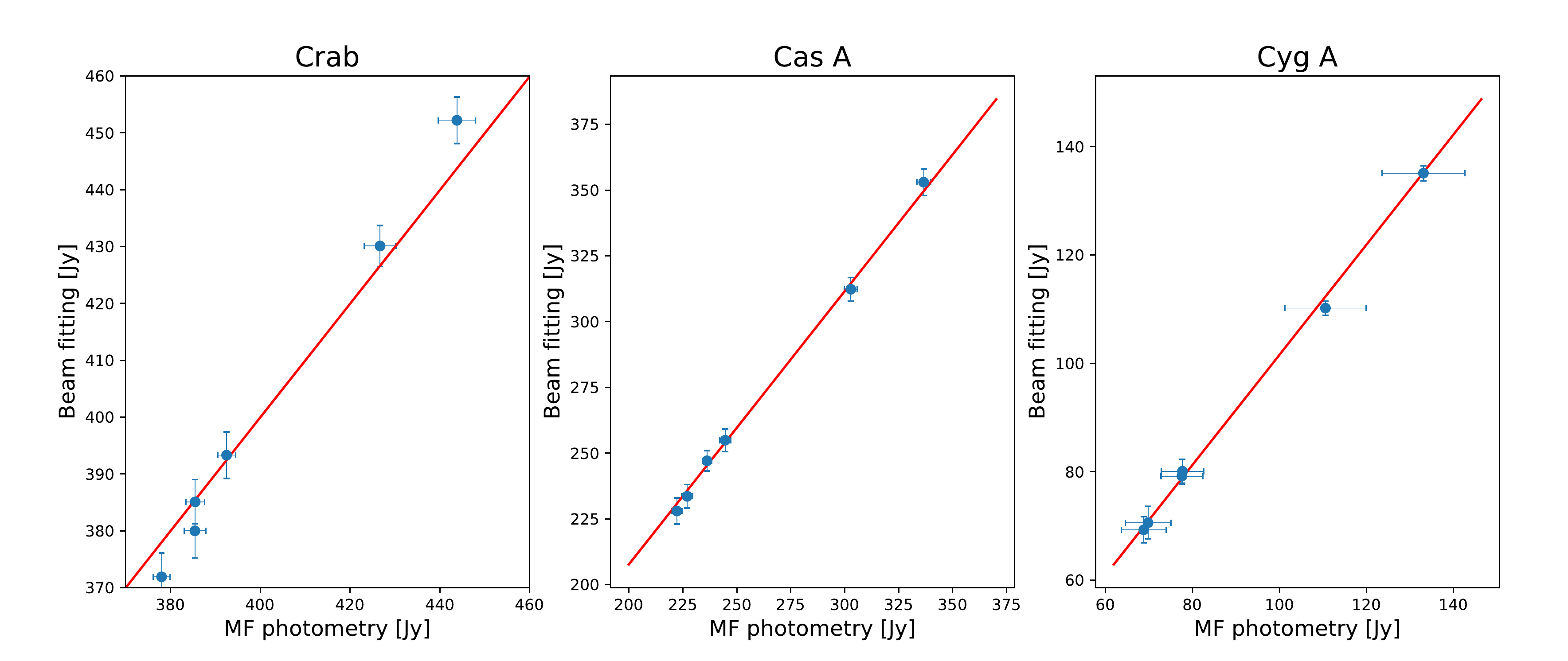}
    \caption{Comparison between the beam-fitting photometry 
used for the Wide Survey map calibration \citep{mfiwidesurvey} 
and the MF photometry used in this paper for the six MFI 
detectors (corresponding to horns 2, 3 and 4) and for the 
three calibration sources Crab, Cassiopeia A and Cygnus A. 
\rtgs{ The red line line shows the best linear $y= a \, x$ fit.}}
    \label{fig:calibrators}
\end{figure*}

Figure~\ref{fig:calibrators} shows the photometric 
measurements used for this paper as compared to 
the prediction from the models that were used to 
calibrate the MFI
real-space beam-fitting photometry in \cite{mfiwidesurvey} 
and \cite{MFIpipeline}. The red line line shows the best 
linear $y= a \, x$ fit. The agreement to a linear law is 
good, particularly for Cas A and Cyg A. 
The linear fit gives slopes of $0.999 \pm 0.005$, 
$1.039 \pm 0.008$ and $1.015 \pm 0.031$ for Crab, Cas A 
and Cyg A respectively. The compelling agreement for the Crab 
is not surprising, as this source has been used as our main 
calibrator \citep{MFIpipeline}. If we compute the percentage 
relative residuals between the two photometries as
\begin{equation}
    r = 100 \left| \frac{I - I_{bf}}{I_{bf}} \right|,
\end{equation}
\noindent 
where $I$ is the flux density estimated in this paper 
and $I_{bf}$ is the beam fitting photometry described in 
\cite{MFIpipeline}, we
get $1.05\%$ mean (from the four frequencies) relative 
errors for the Crab, $3.56 \%$ for Cas A and $1.47 \%$ for 
Cygnus A. The three values are below the $\sim$5 per cent 
calibration uncertainty of the MFI maps.

As an additional consistency test for the MF photometry, we 
have extended the same real-space beam-fitting procedure to 
the whole catalogue.
 We find agreements between the two photometries that are 
roughly similar to the agreement between the MF and  the MHW2 
estimates discussed in the previous section. As an example, at 
11 GHz we find slope $a=1.05$ and intercept $b=-0.25$.

\subsection{VLA observations} \label{sec:observations}

\begin{figure}
	\includegraphics[width=\columnwidth]{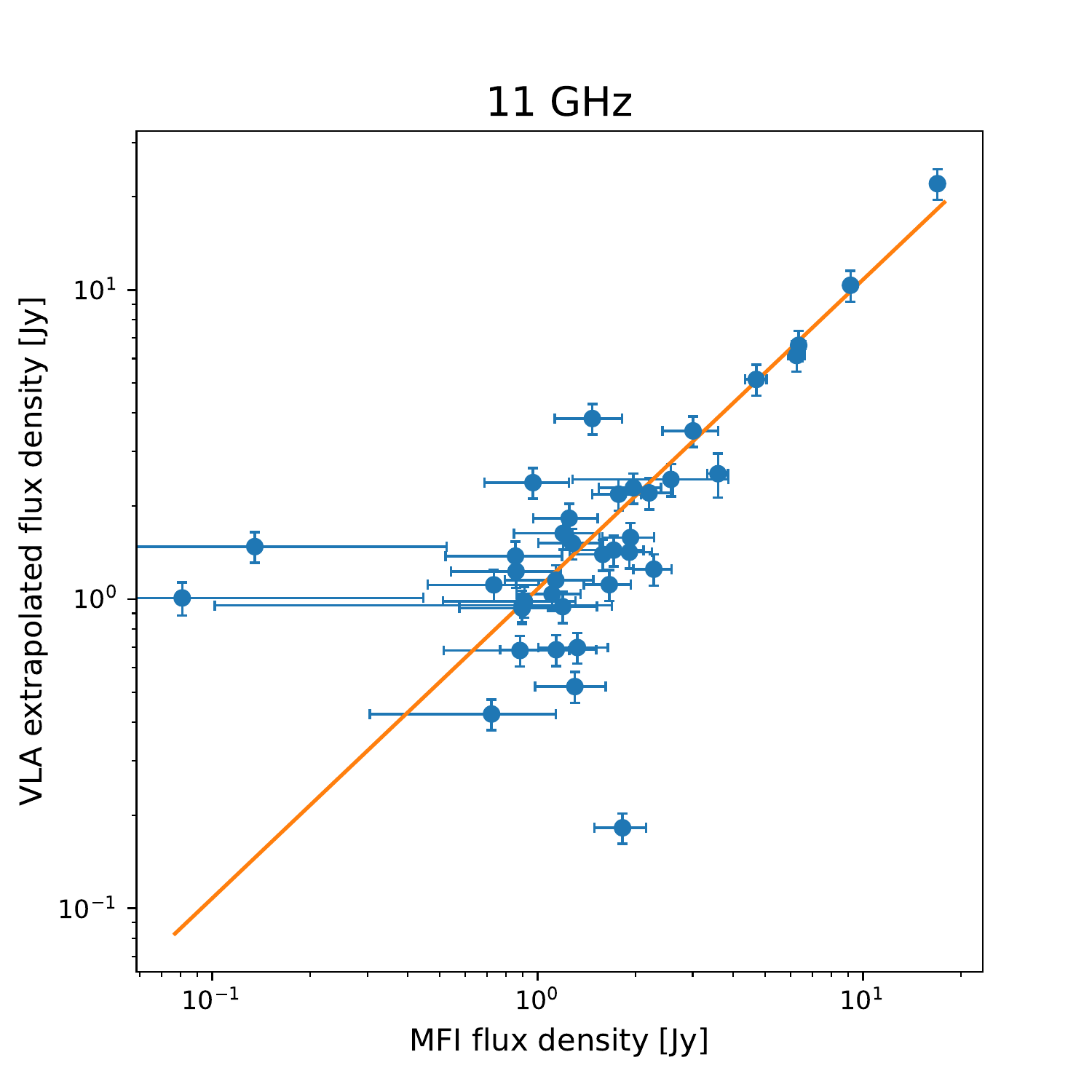}
     \caption{Comparison between the QUIJOTE MFI flux density 
at 11 GHz and the flux density extrapolated from VLA observations 
at 34 GHz for 39 sources in the QUIJOTE cosmological fields 
\citep{VLAsources}. The orange solid line shows the best linear fit.}
    \label{fig:VLA11}
\end{figure}

In parallel to the observation of the QUIJOTE-MFI Wide Survey, 
and as a part of the European Union's Horizon 2020 RADIOFOREGROUNDS 
project,\footnote{\url{http://radioforegrounds.eu/}} \cite{VLAsources} 
observed a sample of 51 sources with the
Very Large Array at 28--40\,GHz. These sources are located
in the QUIJOTE cosmological fields and are
brighter than 1 Jy at 30 GHz in the \textit{Planck} Point 
Source Catalogue. The observations were to characterize their 
high radio-frequency variability and polarization
properties. 
The sources were observed
with a custom correlator configuration that
allowed simultaneous observations at two frequency bands, 28.5--32.4 
and 35.5--39.4 GHz, which were divided into 32 spectral windows, each
with 64 channels of width 2 MHz. Using this configuration, 
\cite{VLAsources}  measured the total intensity flux density 
and spectral index at $\sim$34 GHz of these 51 sources, as 
well as their polarimetric properties (polarization fraction 
and angle, rotation measure).
Since those observations partially overlap in time with the 
QUIJOTE MFI Wide Survey observations, the VLA sample provides 
a good additional testbed for the MFI point source catalogues. 
There are 49 matches (within a 30$^\prime$ search radius) 
between the \cite{VLAsources} VLA sample 
and the sum of our main and extended MFI catalogues.
For this matching, we have  considered only those VLA sources observed
during  the MFI Wide Survey observation campaign (May 2013 to June 2018). 
Five of the sources in our main+extended catalogue 
are resolved or have multiple images in the VLA maps (that is, 
they can be associated with more than one VLA source within the 
30$^\prime$ matching radius). We have not included these sources 
in the following analysis. This leaves 39 matches between our 
catalogue and the VLA sample.
Using the VLA 34 GHz estimated flux density and spectral index, 
we have extrapolated the total intensity flux density of these 
39 matches to MFI frequencies, assuming a single power law 
frequency dependence.  Figure~\ref{fig:VLA11} shows the comparison 
between the MFI (colour-corrected) flux density at 11 GHz 
$S_{\mathrm{MFI}}^{11}$ and the VLA-extrapolated flux density at 
the same frequency  
$S_{\mathrm{VLA}}^{11}$
for the 39 sources in the cosmological field studied by 
\cite{VLAsources} that are in our catalogues. A fit to the law 
$S_{{\scriptscriptstyle 
\mathrm{VLA}}}^{11} =  a \, S_{{\scriptscriptstyle \mathrm{MFI}}}^{11}$ 
gives $a = 1.08 \pm 0.04$.  The agreement is remarkable  considering 
the frequency gap between 11 and 34 GHz, the different 
systematics that affect the VLA and the QUIJOTE MFI, and 
source variability.

\begin{figure}
	\includegraphics[width=\columnwidth]{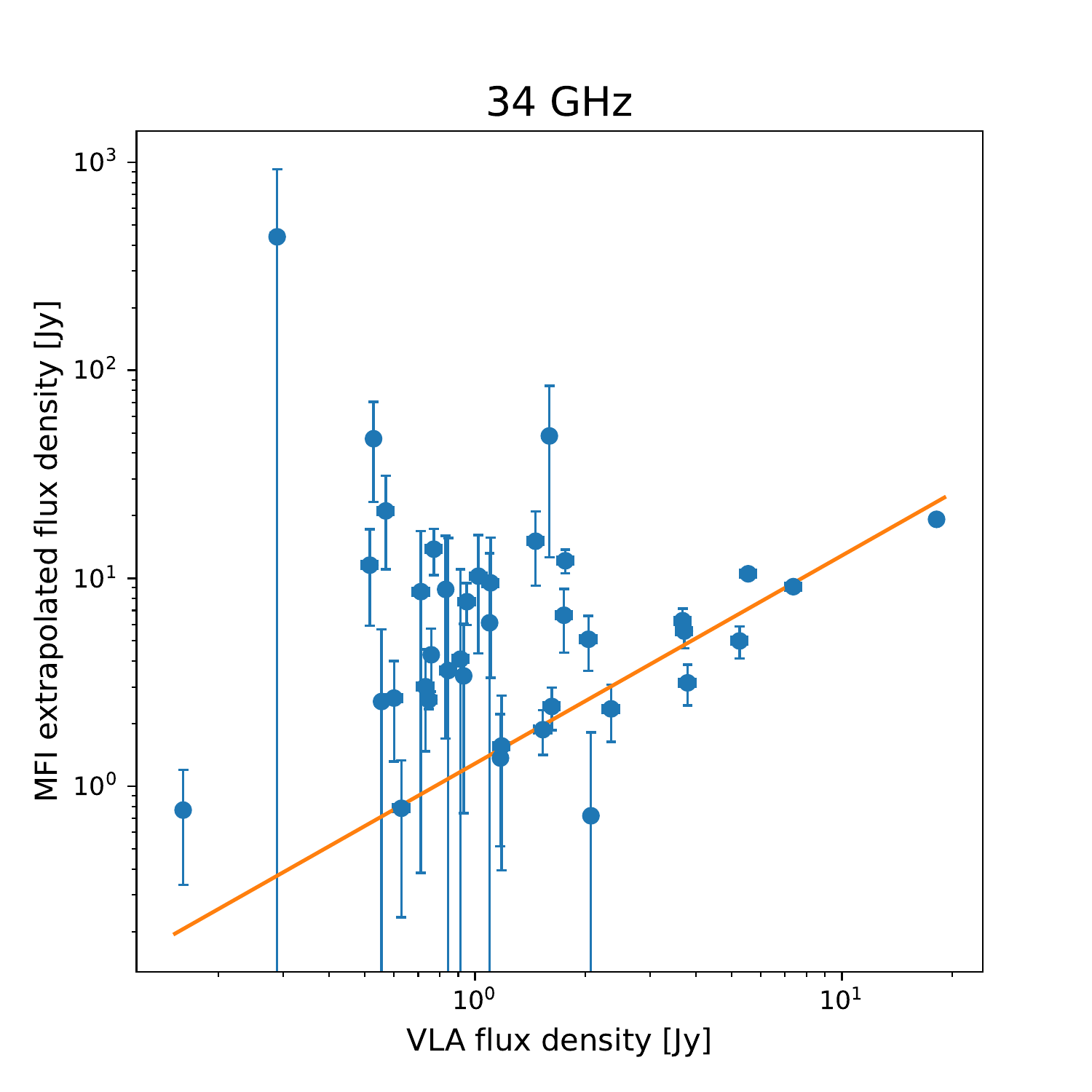}
    \caption{Comparison between VLA observations at 
34\,GHz for 39 sources in the QUIJOTE cosmological fields 
\citep{VLAsources} and the predicted flux at the same 
frequency using spectral index fitted from the QUIJOTE MFI 
 data at 11, 13, 17 and 19 GHz. \chlc{The orange solid 
line shows the best linear fit}.}
    \label{fig:VLA12}
\end{figure}

It is also interesting to repeat the same exercise in the 
opposite way. We have used the MFI flux densities at 
11, 13, 17 and 19 GHz to fit an  intensity and spectral index 
that we have used to predict the flux density of the sources at 
the VLA average frequency of 34 GHz. Figure~\ref{fig:VLA12} 
shows the results of this exercise. MFI predictions tend to 
overestimate the flux density at 34 GHz for sources below 
$\sim$10 Jy. The fit 
$S_{\scriptscriptstyle 
\mathrm{MFI}}^{34} =  b \, S_{{\scriptscriptstyle \mathrm{VLA}}}^{34}$ gives $b = 1.29 \pm 0.04$. 
As expected, the VLA observations give more precise spectral 
index estimates than the MFI data owing to the difference in sensitivity.

\section{Intensity and polarization properties of the sources} \label{sec:analysis}

\subsection{Number counts in total intensity} \label{sec:number_counts}

\begin{figure*}
    \centering
    \includegraphics[width=\textwidth]{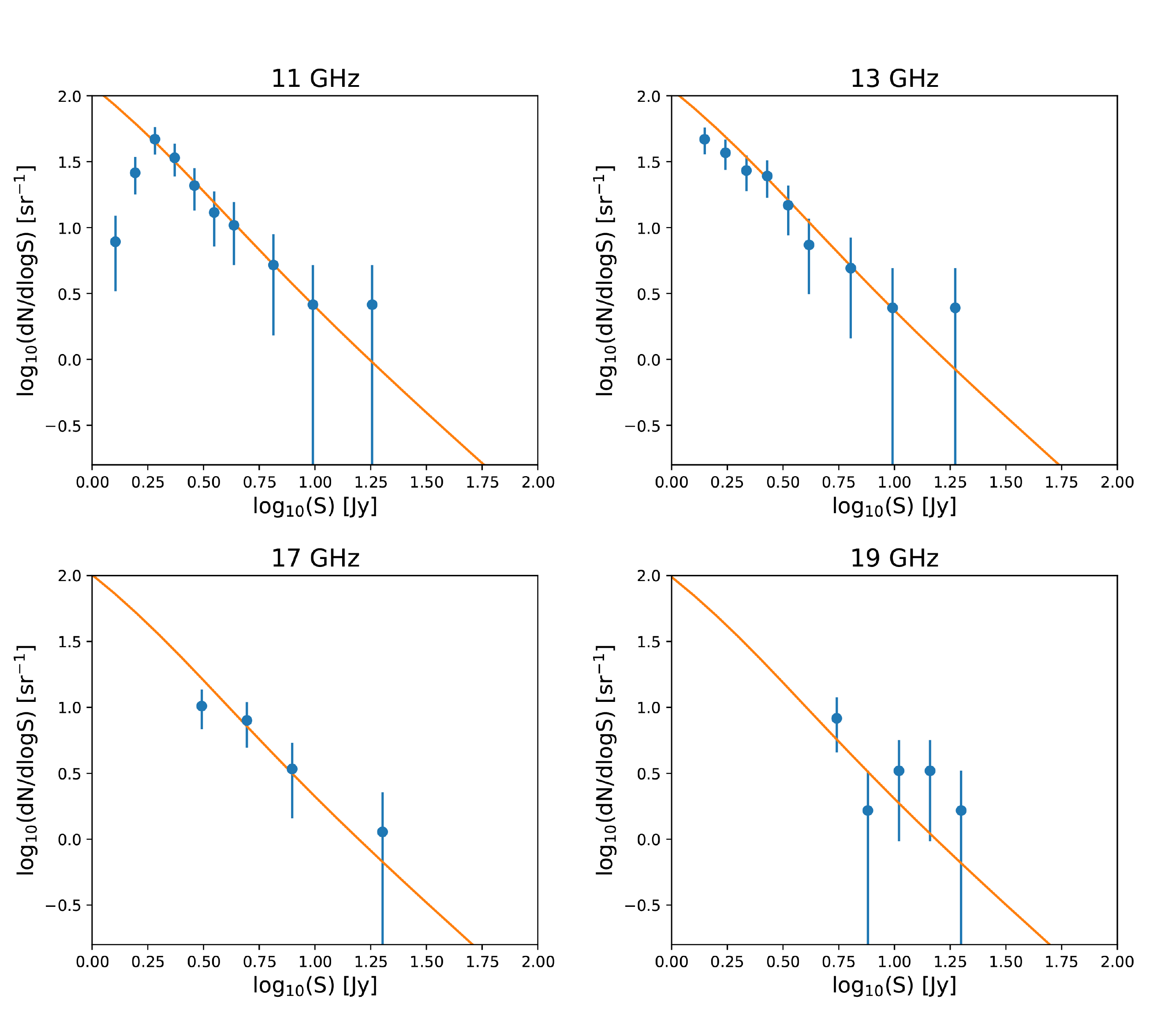}
    \caption{Differential source number counts (blue dots) 
for the MHW2 blind sample at 11, 13, 17 and 19 GHz. Only 
sources with $\protect \sigma \geq 4.5 $ and outside 
    the Galactic mask have been used for this plot. As a 
comparison, the predicted radio source number counts from the
 \protect \cite{zotti05} model are also plotted (continuous orange line). }
    \label{fig:counts}
\end{figure*}

Counts of radio galaxies are one of the traditional ways 
of characterizing the evolutionary properties of this type of 
extragalactic object. However, to study properly  the 
evolution of radio sources we would need to go to flux density 
limits that are well below QUIJOTE's sensitivity. 
Nevertheless, number counts are still interesting as a means 
 of checking the validity of our point source catalogue by comparing 
 its number counts  with well-known models. For this purpose 
the catalogue needs to be blind, 
 so we focus on the MHW2 sources
described in Section~\ref{sec:blind}. 
We use the flux densities obtained by the MHW2, knowing that 
they are consistent with the MF photometry used in the rest 
of the paper (see Section~\ref{sec:MF_vs_MHW2}).  
We have considered only bright 
sources (detected at the $\geq 4.5 \sigma$ level in each channel)
 likely to be extragalactic (outside the \texttt{GAL070} 
 \emph{Planck} Galactic mask\footnote{\texttt{GAL070} is
 one of the \emph{Planck} 2015 Galactic plane masks, with 
no apodization, used for CMB power spectrum estimation. 
The whole set of masks consists of \texttt{GAL020}, 
\texttt{GAL040}, \texttt{GAL060}, \texttt{GAL070}, 
\texttt{GAL080}, \texttt{GAL090}, \texttt{GAL097} and 
\texttt{GAL099}, where the numbers represent the percentage
 of the sky that was left unmasked. 
The \texttt{GAL070} is considered as a safe mask to remove
 most of the Galactic point sources from a catalogue.
These masks can be found online at the Planck Legacy Archive,
 \url{http://pla.esac.esa.int/pla}. See the Planck 
Explanatory Supplement for further description of the 2015 
data release \citep{planck2016-ES}.}).
We have not applied any colour correction to these sources. 
The reason for this decision is that in order to compute 
colour corrections we need to estimate spectral indexes for 
the sources and this can be done only for the intersection 
of the single-frequency catalogues; this would further reduce 
the size of our sample, so we prefer to keep the full blind 
MHW2 catalogues at each frequency,
Figure~\ref{fig:counts} shows the differential number counts 
for the sources satisfying the above conditions at 11, 13, 17 
and 19 GHz (blue dots). The solid orange line shows the number
 count models by \cite{zotti05} at these frequencies (the 
updated counts models by \cite{tucci11} are essentially 
identical to the \cite{zotti05} models for frequencies below 
30 GHz). At 11 and 13 GHz the agreement between the MFI MHW2
 blind catalogue differential source number counts and the 
\cite{zotti05} models is remarkably good and indicates a 
$4.5\sigma$ completeness limit $\sim 1.8$ Jy.
At 17 and 19 GHz  the agreement with the \cite{zotti05} model 
is also reasonably good, but one must take into account that 
the number of $4.5\sigma$ sources outside the Galactic mask 
at these frequencies is too low to give reliable statistics.
 The number of sources used for the plots in Figure~\ref{fig:counts}
 are 67, 70, 21 and 13 for 11, 13, 17 and 19 GHz respectively.

\subsection{Spectral indexes in total intensity}

\begin{figure}
	\includegraphics[width=\columnwidth]{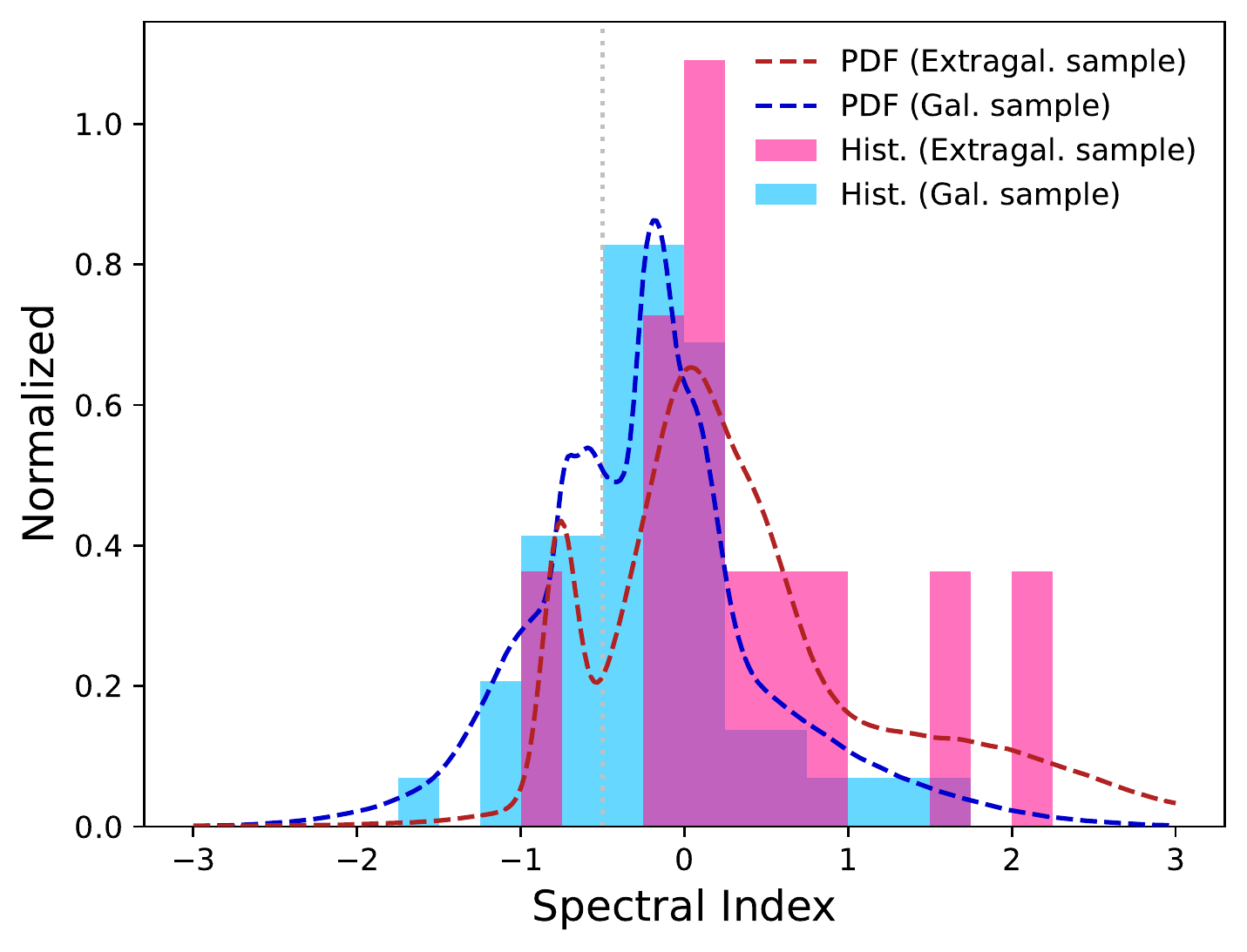}
    \caption{Spectral index distribution of sources in 
the extragalactic (red) and the Galactic (blue) sample.  
    Extragalactic sources are located in the sky region 
observed by the \emph{Planck} \texttt{GAL040} Galactic 
mask (outside the masked area), while the Galactic sources 
are in the complementary area (inside the masked area).
 For each sample, the histogram and probability density 
function (dashed lines) are shown. The area under the 
histograms and the probability density functions integrate 
to 1 (normalized). The grey dotted line establishes the 
$\alpha= -0.5$ limit used to separate flat ($\alpha > -0.5$) 
and steep ($\alpha \leq -0.5$) spectrum sources.}
    \label{fig:spindex}
\end{figure}

\label{sec:spectral_index}

The distribution of spectral indexes
is shown in Figure~\ref{fig:spindex}.
We have considered the spectral indexes only for sources 
detected above the 3$\sigma$ level in all the frequencies 
simultaneously, which amounts to 69 sources. This means 
that only bright sources, $S \geq 1$\,Jy at 11\,GHz, are 
studied in Figure~\ref{fig:spindex}.
In principle, we can distinguish between Galactic and 
extragalatic sources depending on whether they are inside or 
outside the masked area of the \texttt{GAL40} Galactic mask 
respectively. The \texttt{GAL40} mask is more restrictive 
than the \texttt{GAL70} mask used in the previous section 
and therefore more likely to separate Galactic from 
extragalactic sources. We find 58 Galactic and 11 
extragalactic sources, respectively.
In Figure~\ref{fig:spindex}, we present the histogram of 
the spectral indexes for each subsample. 
In addition, we draw the probability density function 
generated as the mean of all marginalized posteriors of 
spectral indexes (dashed lines). 
The extragalactic sample appears to follow a bimodal 
distribution with a majority of flat  radio sources
 \cite[$\alpha \geq -0.5$,][]{zotti05}  and a small portion
 of steep radio sources (only one source). This 
 bimodal distribution of spectral indexes has been reported
 in other experiments at centimetre wavelengths \citep[see,
 for example,][]{PCCS2,PCNT}.
From the integration of the  probability density function 
of the extragalactic sample, between $-\infty$ and $-0.5$, we 
find that $15\%$ are steep-spectrum sources.
This result is near to the $\sim$20\% of steep-spectrum 
radio sources above 1.5\,Jy at 11\,GHz predicted by the 
\cite{zotti05} model, but we must remark that our sample 
size is very limited (only 11 sources satisfy the 
strict criteria we have imposed to be catalogued as 
extragalactic sources). Two sources show spectral indexes 
$\geq 1$. their PCCS2 IDs are PCCS2 030 G002.28+65.92 and
 PCCS2 030 G174.48+69.81 respectively. The first one 
corresponds to the WMAP source WMAP J141552+1324, with no 
known redshift measurement. It is probably associated with
 QSO B1413+135, a blazar at $z=0.2467$.   The second one,
 PCCS2 030 G174.48+69.81, is probably associated with QSO
 B1128+385 at redshift $z=1.7404$. Both seem to 
be highly active, with recent ATels shown in the reference 
list in Simbad.  If the QUIJOTE measurement caught them 
during a flare, the radio emission would be optically thick, 
thus explaining the rising spectrum.

\subsection{Polarimetric properties}

\begin{figure*}
	\includegraphics[width=\textwidth]{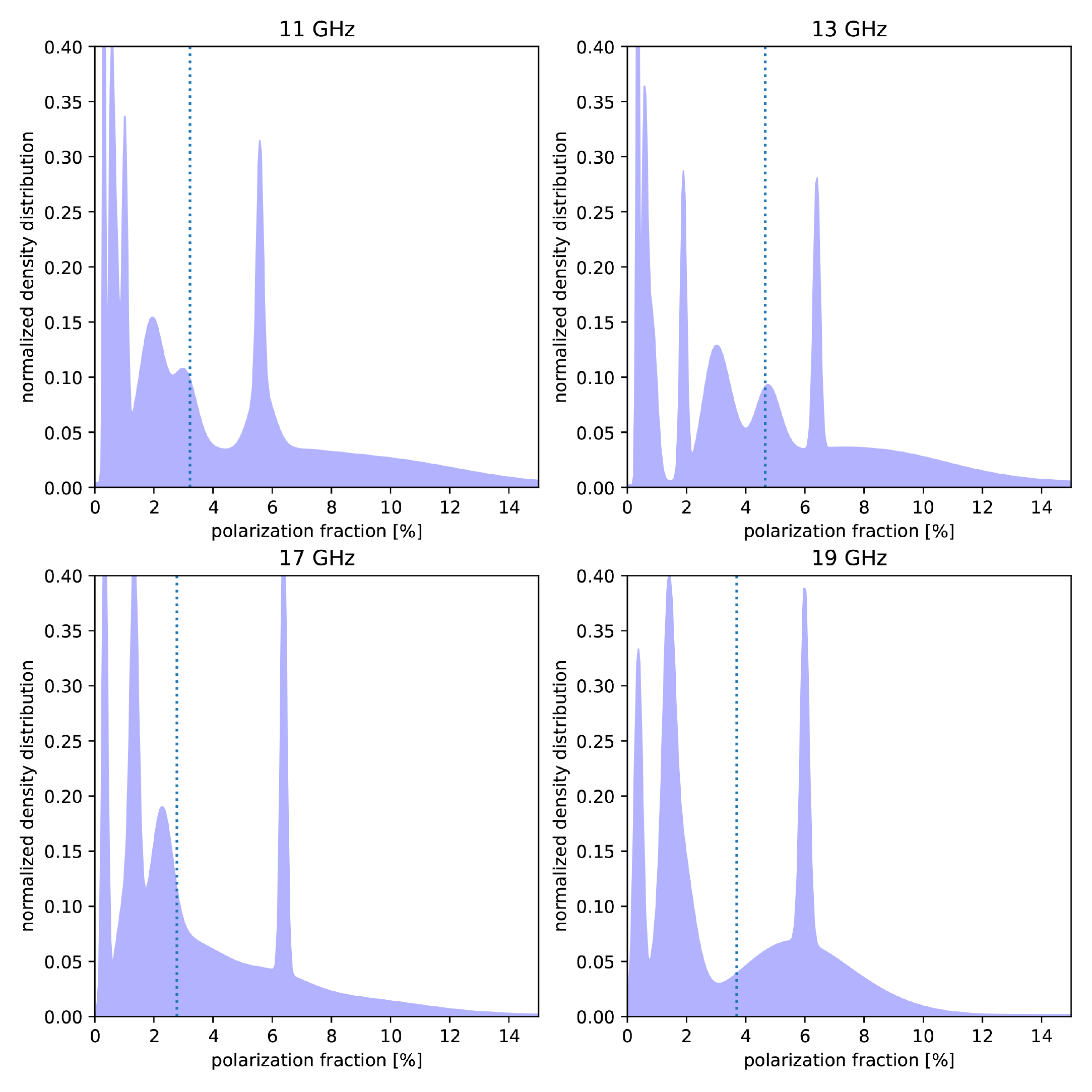}
    \caption{Density distribution of the polarization 
fraction of the $99.99 \%$ s.l.\ sources in the main sample. 
Vertical dotted lines indicate the median polarization 
fraction. The large, thin bump around $\Pi \sim 6\%$ in 
the four panels corresponds to Tau A (the Crab nebula).}
    \label{fig:polfrac}
\end{figure*}

\begin{table}
\begin{centering}
    \begin{tabular}{ccccc}
    \hline
    & 11 GHz & 13 GHz & 17 GHz & 19 GHz \\
    \hline
    N(s.l. $\geq 99.00\%$) &
45 (22) &
46 (27) &
38 (14) &
32 (12) \\
N(s.l. $\geq 99.99\%$) &
38 (18) &
33 (16)&
31 (11)&
23 (9) \\
\hline
    \end{tabular}
    \caption{Number of sources with polarization estimated 
    above a given significance level threshold
    at each of the MFI channels. Numbers between parenthesis indicate the number of
    sources above the same threshold that have Galactic latitude $b\geq 20^\circ$.}
    \label{tab:number_per_Pcl}
    \end{centering}
\end{table}

We have obtained polarization measurements with statistical 
significance level  (s.l.) $\geq 99.99\%$ 
\citep[see][for details]{WMAP_pol}   for 
$(17, 15, 10, 9)$ sources at $(11, 13, 17, 19)$ GHz
 in our main sample. For the extended sample, we have 
found $(21, 18, 21, 14)$ $99.99\%$ s.l.\ sources at 
$(11, 13, 17, 19)$ GHz.  For further reference,
Table~\ref{tab:number_per_Pcl} shows the number of 
sources in the total sample (main plus extended) with 
polarization measurements with statistical significance level 
$\geq 99.00 \%$  and $\geq 99.99 \%$ in the whole 
sky covered by our survey and above $|b|\geq 20^\circ$ 
(in parentheses).
We provide I, Q, P and polarization angle and their 
associated estimated errors  for these sources, as 
well as for  the rest of the catalogue, but we 
remind readers that a high statistical significance 
level of the detection does not necessarily mean a
 small error in the determination of the polarimetric
 parameters. As seen in section~\ref{sec:consistency},
 the design of the MFI 
was not optimized for the study of compact sources, 
and this
makes it difficult to estimate the polarization angle 
of all but the brightest polarized sources. For this 
reason we do not attempt to extract information about 
the rotation measurement (RM) from this catalogue.

\begin{table}
    \centering
    \begin{tabular}{ccc}
    \hline\
      Frequency [GHz]   &  $\Pi_{\mathrm{med}}$ [\%] & $\langle \Pi^2 \rangle$ \\
      \hline
      11 & 3.2 & 0.017 \\
      13 & 4.7 & 0.022 \\
      17 & 2.8 & 0.014 \\
      19 & 3.7 & 0.020 \\
      \hline
    \end{tabular}
    \caption{Median polarization fraction, $\Pi_{\mathrm{med}}$,  and mean value of the squared polarization fraction $\langle \Pi^2 \rangle$ for sources in the main sample with polarization significance level $\geq 99.99 \%$. 
    The values have been calculated from the continuous density functions used in Figure~\ref{fig:polfrac}.
     }
    \label{tab:table6}
\end{table}

We have calculated the polarization fractions of the 
$99.99\%$ s.l.\ sources in our main sample (the 47 
sources with polarization already measured by 
\textit{Planck}). Figure~\ref{fig:polfrac} shows the 
density distribution of polarization fractions for 
these sources, assuming Gaussian 
uncertainties.\footnote{The distribution of errors of 
$\Pi \equiv \hat{P}/\hat{I}$ is not Gaussian, since
 $\hat{P}$ is not normally distributed. However, 
it is reasonable to assume that $\hat{I}$ errors are 
Gaussian, and the Central Limit Theorem indicates
 that errors in $\Pi$ should be somewhat Gaussianised.}
 Vertical lines indicate the median polarization 
fraction, which is $(3.2,4.7,2.8,3.7)\%$ at
 $(11, 13, 17, 19)$ GHz. 
These median values have been computed from the 
cumulative density function (cdf) derived from the 
density distributions used for Figure~\ref{fig:polfrac}. 
The advantage of using the density distributions 
instead of the discrete samples is that in this way 
it is possible easily to take into account the 
uncertainties in the estimation of the polarization fraction.
The values obtained in this work are between median 
values reported in the literature for radio-flat 
sources \citep{sajina11,Puglisi} and  radio-steep 
sources \citep{AT20G,sajina11,Puglisi} below 
$\nu < 20$ GHz.  The subsamples used for the calculation 
of the polarization fraction contain from $14\%$ 
at 13 GHz to $0 \%$ at 17 and 19 GHz radio-steep sources.  
Table~\ref{tab:table6} shows the same median values 
and also the average value of the squared polarization 
fraction $\langle \Pi^2 \rangle$.
These values can be used in combination with the number
 counts to estimate the expected contribution from
radio sources to the polarization power spectra of 
the QUIJOTE MFI data, as described in \cite{lagache20}.
 The expected contribution is
$\la 32,25,11,10$\,$\mu$K.deg at 11, 13, 17 and 19 GHz 
respectively. These values are consistent with the 
predictions of \cite{Puglisi} and the best-fit results 
presented in \cite{mfiwidesurvey}.

The same study for the extended sample gives significantly
 higher median polarization fractions: 
$\Pi_{\mathrm{med}} = (27.3,
  33.6,
  28.7,
  27.0)\%$ at $(11, 13, 17, 19)$ GHz.
  The subsample of the extended catalogue used for the 
calculation of these polarization fractions is clearly 
dominated by steep sources (from a minimum value of $61\%$ 
steep sources found at 13 GHz to a maximum value of $85.7\%$ 
steep sources at 17 GHz).
  The high polarization fractions found for this subsample
 suggest that we may be overestimating the polarized flux 
density of the extended catalogue sources owing either to 
insufficient debiasing (equation~\ref{eq:P_debiased}) or 
to Eddington bias, even for high-significance sources. 

\newpage


\section{Variability study}  \label{sec:variability}
QUIJOTE-MFI data span a period of 6 yr, between November 
2012 and August 2018, which allows for variability studies. 
These data have been separated in six different periods,
 with variable duration (2--22 months), which are calibrated 
independently (see \cite{mfiwidesurvey} and \cite{MFIpipeline} 
for details). The observations leading to the Wide Survey 
maps were taken during periods 1, 2, 5 and 6. \rtgs{Together 
with full-mission maps, maps per individual periods were 
also generated (see details in section~4.1.2 of 
\cite{mfiwidesurvey}). In order to identify variable sources 
(on time scales of $\gtrsim 6$ months), we have computed 
flux densities in total intensity for all sources having 
S/N $>5$ at 11~GHz in the four periods maps.} 

Flux densities extracted from \rtgs{maps of} horns 2 and 4 
are combined using appropriate weights \citep{mfiwidesurvey} 
into one single measurement at both 16.8~GHz and 18.8~GHz. 
\rtgs{A correct estimate of the uncertainties associated 
with our flux density measurements is key to assessing variability.
 This estimate is based on error propagation from the 
scatter of the residuals of fits performed on individual-period 
maps from which we subtract the full map. We proceed in this 
way in order to eliminate the contribution from the sky 
background to the final error bar, which is common in all 
periods. Equally important is the estimate of the `effective 
observing date'.} For each source we calculate this 
effective date as the weighted average of all the dates in 
which the source has \rtgs{been picked up by} the telescope 
main beam (using the same weights that were used to weigh 
all the data lying in the same pixel when producing the final maps).

\begin{figure*}
	\includegraphics[width=\textwidth]{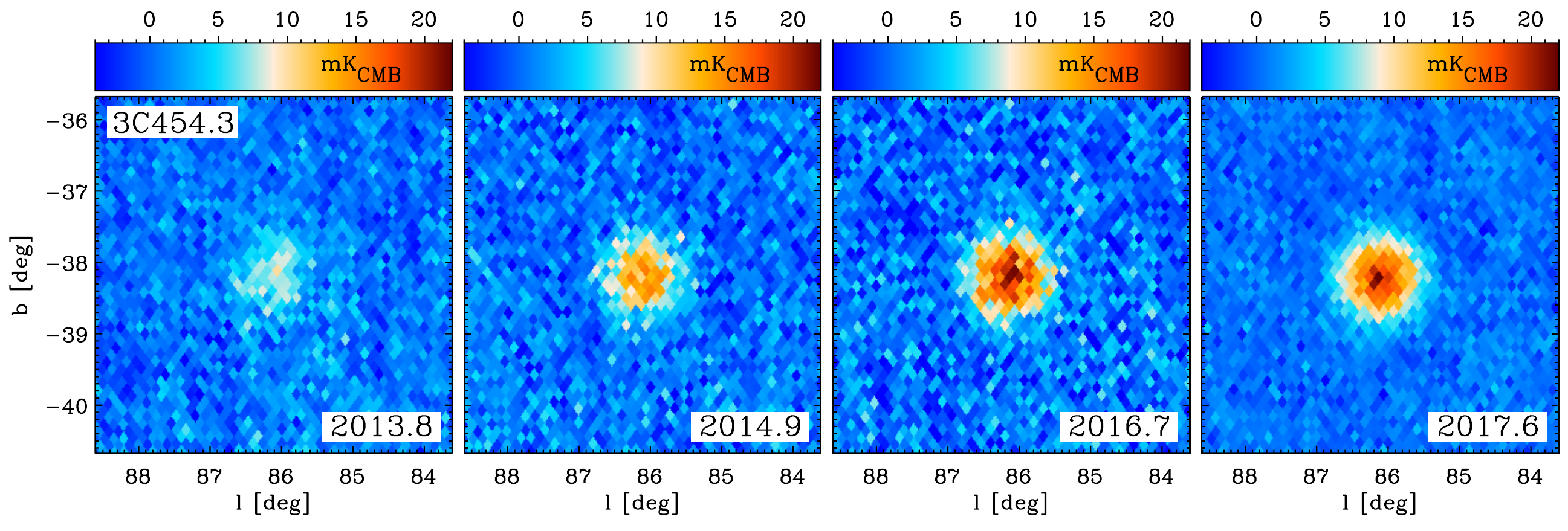}
	\includegraphics[width=\textwidth]{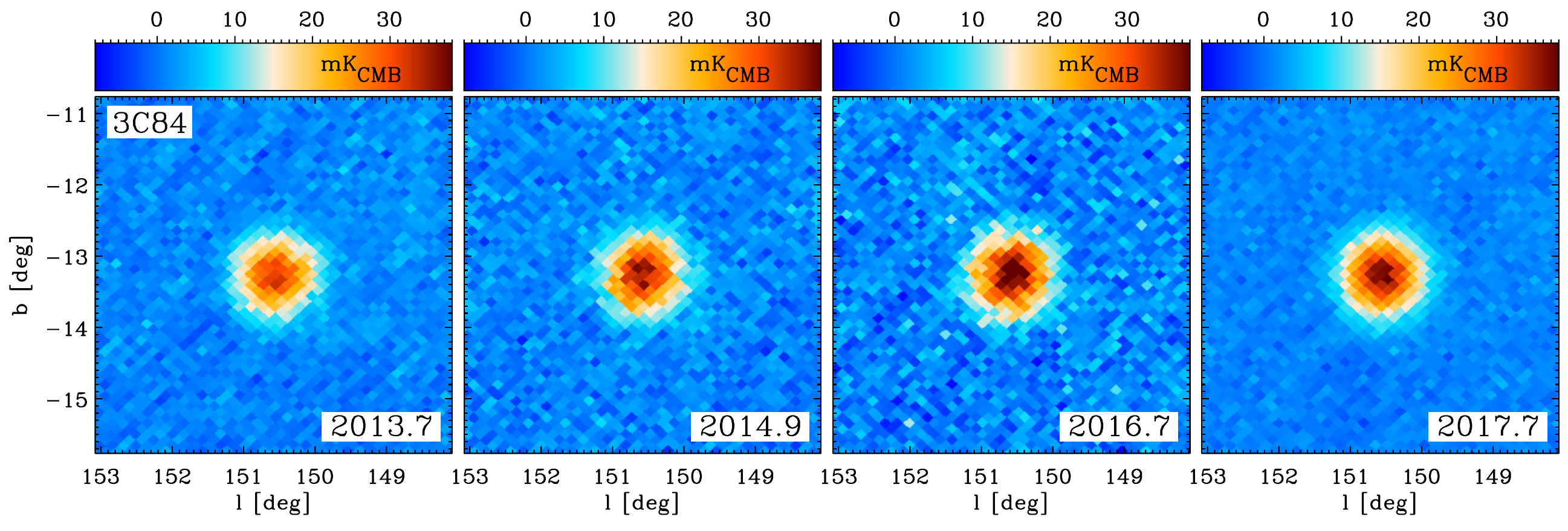}
	\includegraphics[width=\textwidth]{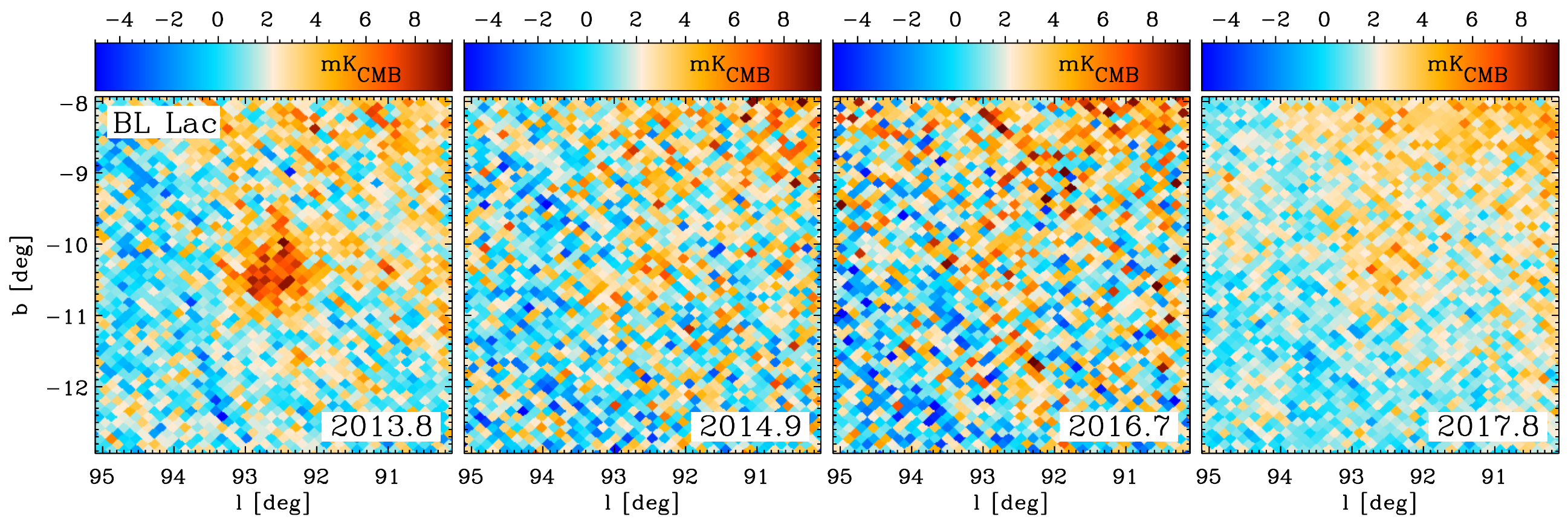}
    \caption{MFI maps at $11.1$~GHz per period (periods 
1, 2, 5 and 6 from left to right) for three of the most 
variable sources identified in the survey: 3C454.3 (top), 
3C84 (middle) and BL Lac (bottom). Effective observing date 
(in years) is shown in the bottom-right corner of each panel.
 For each source the colour scale of the maps has 
intentionally been normalized to the same values for all 
four periods. The change in the brightness of the source is 
appreciable by eye between the different periods.}
    \label{fig:period_maps}
\end{figure*}

\begin{figure*}
	\includegraphics[width=\textwidth]{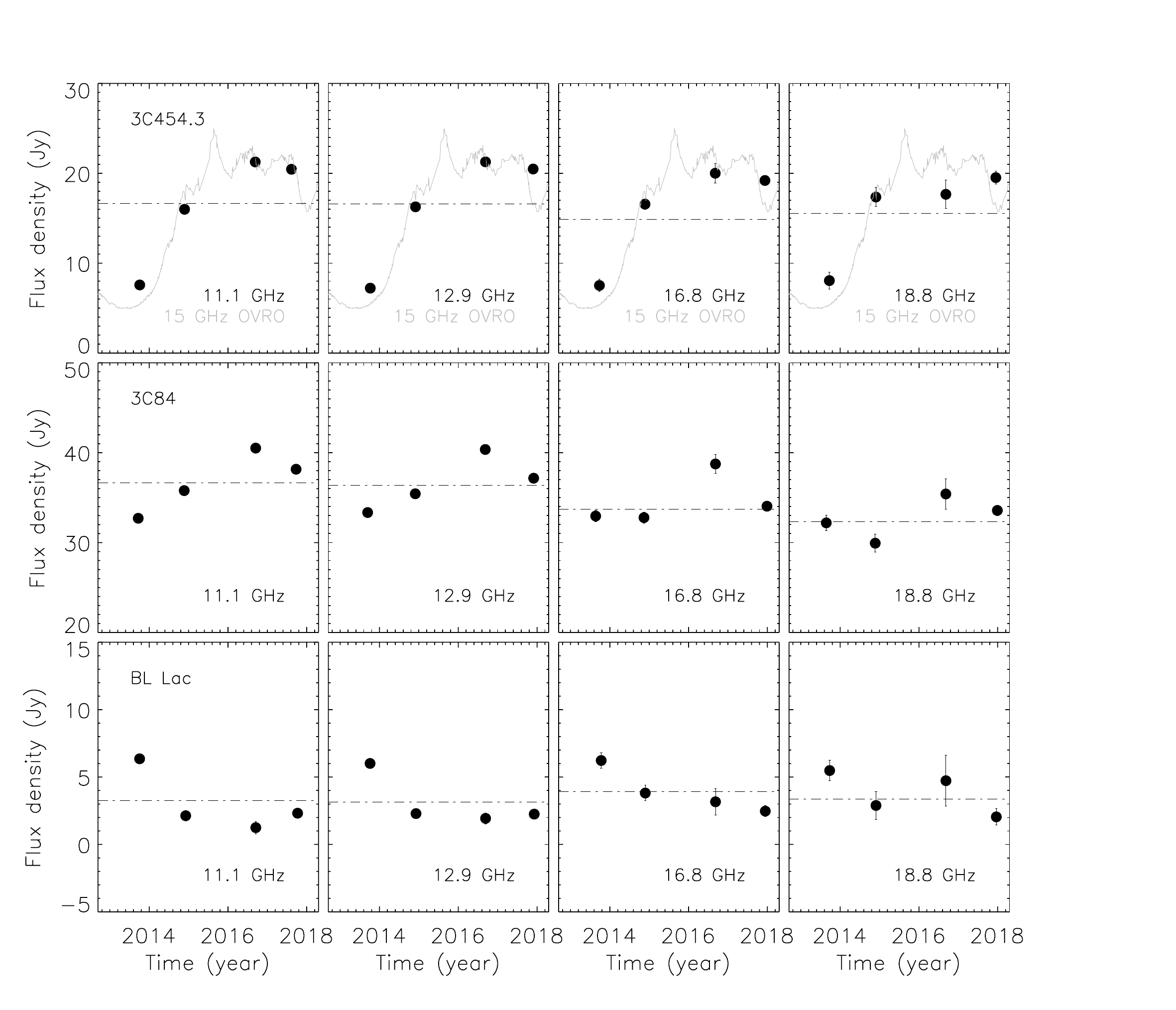}
    \caption{Flux density as a function of time calculated 
on the individual-period maps at the four MFI frequencies (left
 to right) for three of the most variable sources (3C454.3,
 3C84 and BL Lac). The horizontal dot-dashed lines depict  
densities extracted from the full map. For 3C454.3 we have 
overplotted the OVRO 15 GHz measurements.}
    \label{fig:flux_vs_period}
\end{figure*}

Figure~\ref{fig:period_maps} shows \rtgs{individual-period 
maps at $11.1$~GHz} 
for three different sources. Variability can be seen by eye 
in these maps. It is also apparent that the map of period six
 is the most sensitive since it contains more data (22 months).
 For a better visualization of the variability trends, in 
Figure~\ref{fig:flux_vs_period} we plot flux densities versus
 effective observing date for all four frequencies. Similar
 variability trends are observed for the four frequencies, 
despite the lower sensitivity of the two higher-frequency 
bands because of atmospheric contamination. It must be noted 
though that the noise in the two frequency pairs 
(11.1/12.9\,GHz, and 16.8/18.8\,GHz)
is correlated to some extent, so those measurements cannot 
be considered as independent (although the lower- and 
higher-frequency measurements are). These variability trends 
are also similar to those traced by the 37~GHz data of the 
Mets\"ahovi Radio 
Observatory.\footnote{\tt{http://www.metsahovi.fi/AGN/data/}.}

Variability is assessed through a simple $\chi^2$ 
calculation to estimate the significance of the scatter
 of the \rtgs{11~GHz flux densities (we use 11~GHz as a 
reference as it is the most sensitive band)} over the four 
periods with respect to their weighted average \citep{chen13}. 
Table~\ref{tab:variable} shows the seven confirmed variable
 sources presenting the highest $\chi^2$. Note that the 
$\chi^2$ values at high frequencies are typically lower. 
\rtgs{This is} a consequence of these data being 
\rtgs{noisier} because they are more affected by atmospheric 
contamination. With three degrees of freedom, the $99\%$ 
confidence threshold is $\chi^2>11.3$, and all 
seven sources shown in Table~\ref{tab:variable} can therefore be 
considered  as {\it strongly variable} (following the same 
criterion as \citealp{chen13}). \rtgs{These sources} are 
the most compelling ones as they present similar variability
 trends in the four frequencies, which are also similar to 
the variability traced by the Mets\"ahovi data. In total 
we detect 37 sources at 11.1~GHz with $\chi^2>11.3$. At 
lower confidence, we detect hints of variability consistent 
with Mets\"ahovi in five other sources (4C39.25, 
PKS1502+106, 0642+449, 2201+315 and 0133+476). \rtgs{We 
have verified that sources that are found to be non-variable 
have $\chi^2$ per degree-of-freedom  $\sim 1$, which 
bestows confidence on our error bar estimate.}

 \rtgs{This variability study has demonstrated} the 
reliability of the internal gain calibration 
of the QUIJOTE-MFI Wide Survey data, whose accuracy is 
found to be better than $1\%$ \citep{mfiwidesurvey,MFIpipeline}. 
A more detailed and extended study of variability would 
entail binning the data on shorter timescales, which, even 
at the cost of less sensitivity may, reveal strongly 
variable sources over shorter periods. These and other 
studies (such as one on variability in polarization)
 will be presented in a future paper.

\begin{table*}
\begin{tabular}{cccccccc}
\hline
\multicolumn{2}{c}{$11.1$ GHz} & \multicolumn{2}{c}{$12.9$ GHz} & \multicolumn{2}{c}{$16.8$ GHz} & \multicolumn{2}{c}{$18.8$ GHz} \\
Date & $S_\nu^{11}$ (Jy) & Date & $S_\nu^{13}$ (Jy) & Date & $S_\nu^{17}$ (Jy) & Date & $S_\nu^{19}$ (Jy) \\
\hline
\multicolumn{8}{c}{3C454.3}\\
2013.76  &  7.6$\pm$  0.3 &	  2013.77 &   7.2$\pm$  0.3  &    2013.74&    7.5$\pm$  0.7  &    2013.73  &  8.1$\pm$  0.9\\
2014.90  & 16.0$\pm$  0.4 &	  2014.91 &  16.3$\pm$  0.4  &    2014.89&   16.6$\pm$  0.6  &    2014.91  & 17.4$\pm$  1.1\\
2016.69  & 21.3$\pm$  0.5 &	  2016.69 &  21.3$\pm$  0.5  &    2016.68&   20.0$\pm$  1.1  &    2016.68  & 17.7$\pm$  1.6\\
2017.61  & 20.4$\pm$  0.2 &	  2017.90 &  20.5$\pm$  0.2  &    2017.94&   19.2$\pm$  0.5  &    2017.95  & 19.5$\pm$  0.7\\
$\chi^2$ &   1256         &&		   1520              &&  		  205        &&			102	 	  \\
\hline
\multicolumn{8}{c}{3C84}\\
2013.72  & 32.7$\pm$  0.3 &	  2013.70 &  33.3$\pm$  0.3	&   2013.64& 33.0$\pm$  0.6  &    2013.65  & 32.2$\pm$  0.9\\
2014.89  & 35.8$\pm$  0.4 &	  2014.91 &  35.4$\pm$  0.4	&   2014.87& 32.8$\pm$  0.6  &    2014.89  & 29.9$\pm$  1.0\\
2016.70  & 40.5$\pm$  0.6 &	  2016.69 &  40.4$\pm$  0.5	&   2016.68& 38.7$\pm$  1.0  &    2016.68  & 35.4$\pm$  1.7\\
2017.73  & 38.2$\pm$  0.2 &	  2017.91 &  37.2$\pm$  0.2	&   2017.99& 34.0$\pm$  0.4  &    2017.99  & 33.6$\pm$  0.6\\
$\chi^2$ &     352        &&                    216              &&                   27.1     &&                12.9     	   \\
\hline
\multicolumn{8}{c}{Cas A}\\
2013.80  & 356.6$\pm$ 0.4 &	  2013.75 & 315.3$\pm$  0.3	&   2013.83& 249.0$\pm$ 1.0  &    2013.75  &226.3$\pm$  1.1\\
2014.90  & 350.5$\pm$ 0.8 &	  2014.92 & 311.9$\pm$  0.7	&   2014.88& 250.4$\pm$ 0.5  &    2014.91  &232.6$\pm$  0.8\\
2016.70  & 350.4$\pm$ 0.4 &	  2016.69 & 311.3$\pm$  0.3	&   2016.67& 251.2$\pm$ 0.9  &    2016.68  &234.2$\pm$  1.3\\
2017.94  & 348.8$\pm$ 0.3 &	  2018.00 & 308.0$\pm$  0.3	&   2018.11& 243.3$\pm$ 0.7  &    2018.10  &221.0$\pm$  1.0\\
$\chi^2$ &   318          &&                    292              &&                81.3        &&                    102      \\
\hline
\multicolumn{8}{c}{BL Lac}\\
2013.76  &  6.4$\pm$  0.3 &	  2013.76 &  6.0$\pm$  0.3	&   2013.78&  6.2$\pm$ 0.6   &    2013.73  &  5.5$\pm$ 0.8\\
2014.93  &  2.1$\pm$  0.4 &	  2014.93 &  2.3$\pm$  0.4	&   2014.90&  3.8$\pm$ 0.6   &    2014.91  &  2.9$\pm$ 1.0\\
2016.71  &  1.2$\pm$  0.5 &	  2016.69 &  1.9$\pm$  0.4	&   2016.68&  3.2$\pm$ 1.0   &    2016.69  &  4.7$\pm$ 1.9\\
2017.77  &  2.3$\pm$  0.2 &	  2017.93 &  2.2$\pm$  0.1	&   2017.95&  2.5$\pm$ 0.4   &    2017.97  &  2.0$\pm$ 0.6\\
$\chi^2$ &     196        &&                    168              &&              29.6          &&                 13.1    	  \\
\hline
\multicolumn{8}{c}{OJ287}\\
2013.71  &  3.5$\pm$  0.3 &	  2013.67  &  3.7$\pm$  0.3	 &   2013.64&  3.1$\pm$ 0.8   &      2013.64&  2.1$\pm$ 1.9\\
2014.86  &  5.0$\pm$  0.4 &	  2014.89  &  5.2$\pm$  0.4	 &   2014.78&  5.1$\pm$ 0.6   &      2014.85&  4.1$\pm$ 1.2\\
2016.69  &  8.3$\pm$  0.5 &	  2016.70  &  7.7$\pm$  0.6	 &   2016.69&  6.9$\pm$ 1.3   &      2016.69&  5.7$\pm$ 2.0\\
2017.91  &  7.4$\pm$  0.2 &	  2017.98  &  8.0$\pm$  0.2	 &   2018.00&  8.4$\pm$ 0.5   &      2018.00&  8.5$\pm$ 0.7\\
$\chi^2$ &   127          &&                    137               &&                40.6        &&                 18.1       \\
\hline
\multicolumn{8}{c}{Tau A}\\
2013.63  &454.0$\pm$  0.6 &	  2013.62  &432.3$\pm$  0.6	 &  2013.54 & 391.3$\pm$ 1.0  &      2013.56&377.7$\pm$ 1.3\\
2014.88  &451.5$\pm$  0.9 &	  2014.91  &429.8$\pm$  0.8	 &  2014.86 & 380.8$\pm$ 0.9  &      2014.89&367.8$\pm$ 1.2\\
2016.70  &450.4$\pm$  0.7 &	  2016.69  &428.1$\pm$  0.7	 &  2016.69 & 386.9$\pm$ 1.4  &      2016.69&373.8$\pm$ 2.0\\
2017.84  &449.6$\pm$  0.4 &	  2017.92  &427.2$\pm$  0.4	 &  2018.09 & 384.1$\pm$ 0.6  &      2018.08&371.2$\pm$ 0.7\\
$\chi^2$ &   40.8         &&                     63.4             &&                 69.3       &&                  34.9      \\
\hline
\multicolumn{8}{c}{PKS1749+096}\\
2013.77  &  3.5$\pm$  0.3 &	  2013.76  &  3.1$\pm$  0.3	 &  2013.85 &	2.2$\pm$  0.6 &      2013.74&  0.7$\pm$ 0.8\\
2014.90  &  5.0$\pm$  0.4 &	  2014.90  &  4.9$\pm$  0.4	 &  2014.87 &	4.2$\pm$  0.6 &      2014.90&  3.4$\pm$ 1.1\\
2016.69  &  3.4$\pm$  0.5 &	  2016.69  &  3.6$\pm$  0.5	 &  2016.70 &	3.8$\pm$  1.2 &      2016.70&  1.4$\pm$ 1.9\\
2017.67  &  2.2$\pm$  0.2 &	  2017.97  &  2.8$\pm$  0.2	 &  2018.04 &	3.2$\pm$  0.3 &      2018.04&  4.0$\pm$ 0.5\\
$\chi^2$ &	 36.6     &&                   26.0               &&                5.7         &&                  13.7 	   \\
\hline
\end{tabular}
\caption{Flux densities and effective observation dates of variable sources identified in the QUIJOTE-MFI survey. We show flux densities at the four MFI frequencies derived from maps of the four different periods, and the effective observation date (year). We also show $\chi^2$ values representing the scatter around the average value. The first three sources (3C454.3, 3C84, Cas A) are identified as strongly variable ($\chi^2>11.3$, with 3 degrees of freedom, corresponding to confidence $>99\%$). The other four, despite having a lower confidence, are known to be variable and show variability trends that are compatible with external datasets at similar frequencies.}
\label{tab:variable}
\end{table*}

\section{Discussion and conclusions} \label{sec:discussion}

In this paper, part of a series describing the data 
analysis and scientific results from the QUIJOTE MFI 
Wide Survey, we have studied 786 compact source 
candidates in the 11--19 GHz frequency range in both 
intensity and polarization. We have divided our sample 
into two catalogues: a main subsample containing 47 bright 
radio sources whose polarization has been measured by 
the \emph{Planck} satellite \citep{PCCS2} and an extended 
subsample containing both 725 targets 
selected with flux density $\geq 1$ Jy at 30 GHz from
 the Second \emph{Planck} Catalogue of Compact  Sources \citep{PCCS2}
and 14 additional SNR $\geq 4$ targets found by means
 of  a blind search performed with the Mexican Hat 
Wavelet 2. In total, we have studied 786 targets which 
are being made public through the RADIOFOREGROUNDS web site.  

The sources have been studied using a Python version 
of \texttt{IFCAPOL}, a software tool that implements a 
matched filter for the study of intensity and the 
Filtered Fusion \citep[FF,][]{FFpaper,WMAP_pol}  for 
the estimation of the polarimetric properties of the 
sources. For each of the targets we have estimated 
the (colour-corrected) I, Q and U Stokes parameters,  
the derived P  and polarization angle measurements, 
and their associated uncertainties. We have also 
determined the statistical significance of the 
detection of the polarized signal as described in 
\cite{WMAP_pol}. For those sources with a clear 
detection, we have estimated the spectral index 
between 11 and 19 GHz (assuming a simple power-law
 emission) in both intensity and polarization. 

We have performed a number of internal and external 
consistency tests. The internal consistency tests, 
using half-mission maps, have served to test both 
the stability of the instrument and the reliability 
of our photometry. We have also focused on three 
bright sources that have been used as calibrators 
(the Crab nebula, Cygnus A and Cassiopeia A) and 
checked that the photometry used in this paper is 
consistent with other photometric estimators (namely, 
aperture photometry and beam fitting) within the 
calibration uncertainty limit of the MFI maps. As 
an external consistency check, we have compared the 
photometry of 39 sources in our catalogues that have 
been observed during the same epoch with the VLA at 
$\sim$34 GHz. Though the comparison is  difficult due to 
 the frequency difference, the need to extrapolate 
using a simple power law, the difference in resolution 
and the variability of the sources, we find
an overall good agreement between VLA and MFI 
observations. As an additional test, we have also 
studied the variability of a selected sample of 
sources in the MFI data and found it to be consistent with 
Mets\"ahovi Radio Observatory observations.

We have also studied the statistical properties of 
our catalogue. A significant fraction of the sample 
is dominated by sources that are probably Galactic: 
177 out of 786 (22.5\%) sources have Galactic 
latitude $|b|\leq 10^\circ$. The number rises to 
307 out of 786 (39.1\%) if we allow $|b|\leq 20^\circ$.
In spite of this, we have computed number counts 
(in total intensity only) with the  sources that are 
likely to be extragalactic. Our results show good 
agreement with the \cite{zotti05} model and suggest 
that our catalogue has a completeness limit at the 
$4.5\sigma$ level $\sim 1.8$ Jy at 11 GHz. 
We have also studied the distribution of colour-corrected 
spectral  indexes in temperature for both Galactic 
and extragalactic sources. We find 15.2 per cent of 
steep sources outside the Galactic mask we have used, 
roughly consistent with the predictions of the \cite{zotti05} 
model, but we must note that our statistics is very poor. 

Finally, we have studied the polarimetric properties of 
(38, 33, 31, 23) sources with P detected with statistical 
significance $\geq 99.99 \%$ at (11, 13, 17, 19) GHz 
respectively. MFI noise levels make it difficult to 
estimate the polarization angle of all but the brightest 
polarized sources. For this reason we do not attempt to 
extract information about the rotation measurement (RM) 
from this catalogue. For our main sample, we find median 
polarization fractions of 
$(3.2,4.7,2.8,3.7)\%$ at $(11, 13, 17, 19)$ GHz. 
These values are between median values reported in the 
literature for radio flat sources \citep{sajina11,Puglisi} 
and  radio steep sources \citep{AT20G,sajina11,Puglisi} 
below $\nu < 20$ GHz. The median polarization fraction we 
find in our extended sample exceeds these values, which 
suggests that we may be overestimating the polarized flux 
density of the extended catalogue sources.

\section*{Acknowledgements}

We thank the staff of the Teide Observatory for invaluable assistance in the commissioning and operation of QUIJOTE.
The {\it QUIJOTE} experiment is being developed by the Instituto de Astrof\'\i sica de Canarias (IAC),
the Instituto de F\'\i sica de Cantabria (IFCA), and the Universities of Cantabria, Manchester and Cambridge.
Partial financial support was provided by the Spanish Ministry of Science and Innovation 
under the projects AYA2007-68058-C03-01, AYA2007-68058-C03-02,
AYA2010-21766-C03-01, AYA2010-21766-C03-02, AYA2014-60438-P,
ESP2015-70646-C2-1-R, AYA2017-84185-P, ESP2017-83921-C2-1-R,
AYA2017-90675-REDC (co-funded with EU FEDER funds),
PGC2018-101814-B-I00, 
PID2019-110610RB-C21, PID2020-120514GB-I00, IACA13-3E-2336, IACA15-BE-3707, EQC2018-004918-P, the Severo Ochoa Programs SEV-2015-0548 and CEX2019-000920-S, the
Mar\'\i a de Maeztu Program MDM-2017-0765, and by the Consolider-Ingenio project CSD2010-00064 (EPI: Exploring
the Physics of Inflation). 
DT acknowledges the support from the Chinese Academy of Sciences (CAS) President's International Fellowship Initiative (PIFI) with Grant N. 2020PM0042.
FP acknowledges support from the Spanish State Research Agency (AEI) under grant number PID2019-105552RB-C43. 
We acknowledge support from the ACIISI, Consejer\'\i a de Econom\'\i a, Conocimiento y 
Empleo del Gobierno de Canarias and the European Regional Development Fund (ERDF) under grant with reference ProID2020010108.
This project has received funding from the European Union's Horizon 2020 research and innovation program under
grant agreement number 687312 (RADIOFOREGROUNDS).

This research has made use of data from the OVRO 40 m 
monitoring program \citep{OVRO}, supported by private 
funding from the California Insitute of Technology and 
the Max Planck Institute for Radio Astronomy, and by NASA 
grants NNX08AW31G, NNX11A043G, and NNX14AQ89G and NSF 
grants AST-0808050 and AST-1109911.

Some of the results in this paper have been derived using 
the \texttt{healpy} and \texttt{HEALPix} packages \citep{healpix,healpy}. 
The packages
\texttt{astropy} \citep{astropy1, astropy2}, 
\texttt{scipy} \citep{scipy},
\texttt{matplotlib} \citep{matplotlib}, \texttt{numpy} 
\citep{numpy} and \texttt{emcee} \citep{2013PASP..125..306F} 
have been extensively used for data analysis and plotting.

\section*{Data availability}

All data products can be freely downloaded from the 
QUIJOTE web, \footnote{\url{http://research.iac.es/proyecto/cmb/quijote}} 
as well as from the RADIOFOREGROUNDS 
platform.\footnote{\url{http://www.radioforegrounds.eu/}}
 They include also an Explanatory Supplement describing 
the data formats. Maps will be submitted also to the Planck 
Legacy Archive (PLA) interface and the LAMBDA site. 
Any other derived data products described in this paper 
(half-survey maps, simulations, etc.) are available upon 
request to the QUIJOTE collaboration.



\bibliographystyle{mnras}
\bibliography{Bib/paper,Bib/soft,Bib/quijote} 




\bsp	
\label{lastpage}

\end{document}